\RequirePackage{fix-cm}
\documentclass[smallcondensed]{svjour3-preprint}     % onecolumn (ditto)
\smartqed  % flush right qed marks, e.g. at end of proof
\usepackage{graphicx}
\usepackage{amsmath}    % for general math symbols
\usepackage{mathtools}  % for \coloneqq
\usepackage{marginnote} % for comments on the page margin

\newcommand{\dv}{$\Delta v$}
\newcommand{\musun}{\mu_{\text{Sun}}}
\newcommand{\muearth}{\mu_{\text{Earth}}}

\journalname{The Journal of the Astronautical Sciences}
\begin{document}

\title{Trajectory Design for the ESA LISA Mission%\thanks{Grants or other notes
}

\titlerunning{Trajectory Design for the ESA LISA Mission}        % if too long for running head

\author{Waldemar Martens         \and
        Eric Joffre %etc.
}


\institute{W. Martens \at
           European Space Operation Centre, Robert-Bosch-Str. 5, 64293 Darmstadt, Germany \\
              Tel.: +49 6151 90 2217\\
              \email{waldemar.martens@esa.com}           %  \\
           \and
           E. Joffre \at
              European Space Research \& Technology Centre, Keplerlaan 1, 2200 AG Noordwijk, Netherlands\\
							Tel.: +31 (0) 71 565 6702\\
              \email{eric.joffre@esa.com}           %  \\
}

\date{ }

\maketitle

\begin{abstract}
The three Laser Interferometer Space Antenna (LISA) spacecraft are going to be placed in a triangular formation in an Earth-trailing or Earth-leading orbit. They will be launched together on a single rocket and transferred to that science orbit using Solar Electric Propulsion. Since the transfer~\dv~depends on the chosen science orbit, both transfer and science orbit have been optimised together. For a thrust level of 90 mN, an allocation of 1092 m/s per spacecraft is sufficient for an all-year launch in 2034. For every launch month a dedicated science orbit is designed with a corner angle variation of close to $60^\circ \pm 1.0^\circ$ and an arm length rate of maximum 10 m/s. Moreover, a detailed navigation analysis of the science orbit insertion and the impact on insertion errors on the constellation stability has been conducted. The analysis shows that Range/Doppler measurements together with a series of correction manoeuvres at the beginning of the science orbit phase can reduce insertion dispersions to a level where corner angle variations remain at about $60^\circ \pm 1.1^\circ$ at $99\%$ C.L.. However, the situation can become significantly worse if the self-gravity accelerations acting during the science orbit phase are not sufficiently characterised prior to science orbit insertion.

\end{abstract}

%%%%%%%%%%%%%%%%%%%%%%%%%%%%%%%%%%%%%%%%%%%%%%%%%%%%%%%%%%%%%%%%%%%
\section{Introduction}
\label{sec:intro}
%%%%%%%%%%%%%%%%%%%%%%%%%%%%%%%%%%%%%%%%%%%%%%%%%%%%%%%%%%%%%%%%%%%
The LISA mission has been selected as the L3 cornerstone mission by the European Space Agency (ESA) in June 2017~\cite{DanzmannLisaProposal,GravitationalUniverse}. It will be the first space-based Gravitational Wave detector and will explore a new frequency band and therefore be sensitive to previously undiscovered sources of Gravitational Waves~\cite{LIGOsensitivity}. LISA will be based on laser interferometry between free-flying test masses inside drag-free spacecraft. In order to realise that concept, the three LISA spacecraft will be placed into a heliocentric Earth-trailing or Earth-leading orbit. Together they will form the corners of a nearly equilateral triangular formation with $2.5\times 10^6$ km arm length~\cite{BenderLISA}: the so-called cartwheel formation. Each spacecraft is forced to follow its two test masses along each of the two laser beam axes they define. The orbits have to be selected such that the stability of the constellation for the planned mission lifetime of 10 years (6 years nominal + 4 years extension) is guaranteed. The main stability requirements are currently.
\begin{enumerate}
	\item The corner angle variations during 10 years shall not exceed $60^\circ \pm 1.0^\circ$. This constraint comes mainly from the limitation of the optical assembly tracking mechanism (OATM) which ensures that the laser beams point into the correct direction at all times.
	\item The relative velocity (arm length rate) shall not exceed 10 m/s during 10 years mission duration. This is due to the bandwidth limitation of the interferometric beat note detection of the system~\cite{Phasemeter}.
	\item The arm length shall be constrained within $2.5\times 10^6 \text{ km } \pm 2.5\times 10^5\text{ km}$ during 10 years. In practice this constraint is never limiting, however.
\end{enumerate}
These stability requirements can be achieved by maximising the distance of the formation to Earth, and thus minimising its gravitational impact on the formation. On the other hand, communication requirements place a limit on the maximum Earth distance, which is currently assumed to be $65\times 10^6$ km. This value, however, depends on the design of the spacecraft communications system.

LISA is going to be launched on a single Ariane 64 rocket and transferred to its operational orbit using Solar Electric Propulsion (SEP). SEP has the advantage of a more efficient use of propellant mass and thus a higher dry mass fraction compared to chemical propulsion. Previous trajectory analyses for LISA~\cite{HughesOrbit,FundamentalsStableFlight,MinimumFlexing,PovoleriKemble,LisaRendezvous,LisaEarthMoon,JoffreICATT} have mainly focussed on the optimal cartwheel orbit design and less on the transfer. Moreover, the interdependency between the transfer and the optimal cartwheel orbit has not been treated before. The present paper is going to discuss the cartwheel orbit design at different levels of accuracy: Starting from a review of analytical models to finally presenting fully numerical results for both optimal transfer and cartwheel orbits.

The question of stability of the cartwheel orbit under insertion uncertainties has not been analysed before to the authors' knowledge. This is going to be addressed in depth in the second part of the present paper. 

The document is structured as follows: after a summary of the most important analytical cartwheel models and results in chapter~\ref{sec:cartwheelAnalytical}, a fully numerical simulation of the optimal cartwheel orbit is presented in chapter~\ref{sec:cartwheelNumerical}. Chapter~\ref{sec:SEPtransfer} will treat the optimisation of the SEP transfer and the interdependency between the transfer and the cartwheel orbit design. A full navigation analysis taking into account all major sources of uncertainty will be presented in chapter~\ref{sec:navigation}. Also, the impact of the computed cartwheel insertion accuracy on the stability of the orbit during 10 years of science phase will be addressed there. Finally, chapter~\ref{sec:conclusions} will summarize the main conclusions.

%%%%%%%%%%%%%%%%%%%%%%%%%%%%%%%%%%%%%%%%%%%%%%%%%%%%%%%%%%%%%%%%%%%
\section{The cartwheel formation - analytic models}
\label{sec:cartwheelAnalytical}
%%%%%%%%%%%%%%%%%%%%%%%%%%%%%%%%%%%%%%%%%%%%%%%%%%%%%%%%%%%%%%%%%%%
In the following a series of analytic models of the cartwheel formation with increasing level of fidelity will be discussed.

\subsection{Linear two-body model}
\label{subsec:linearTwoBody}
\begin{figure}
  \includegraphics[width=0.5\textwidth]{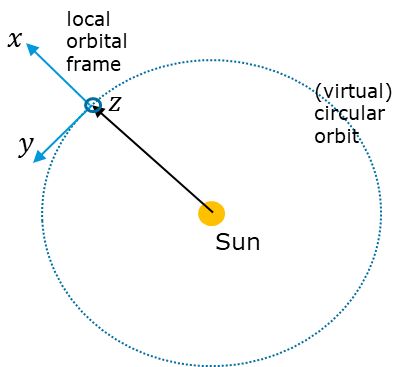}
\caption{Illustration of the local orbital frame.}
\label{fig:localOrbitalFrame}     
\end{figure}

In the two-body problem, the linearised relative motion around a centre on a circular orbit is described by the Clohessy-Wiltshire equations~\cite{ClohessyWiltshire}. Imposing no along-track drift, these equations have the solution~\cite{SpacecraftFormationFlying,Amico2006}:
\begin{align}
\label{eq:linearTwoBody}
	x(t) &= \varrho_x \sin\left(\frac{2 \pi}{T} t + \alpha_x\right) \, ,\\
	y(t) &= \varrho_y + 2 \varrho_x \cos\left(\frac{2 \pi}{T} t + \alpha_x\right) \, ,\\
	z(t) &= \varrho_z \sin\left(\frac{2 \pi}{T} t + \alpha_z\right) \, ,
\end{align}
where $x(t) ,y(t), z(t)$ are the components of the local orbital frame of the virtual formation centre as a function of time:
\begin{itemize}
	\item $x = $ radial (oriented positively from central body to spacecraft)
	\item $z = $ cross-track (along spacecraft angular momentum vector)
	\item $y =$ completing the right-handed frame (along-track for a circular spacecraft orbit)
\end{itemize}
This is illustrated in Figure~\ref{fig:localOrbitalFrame}. $T$ is the orbital period of the circular formation centre orbit and $\alpha_i$ and $\varrho_i$ are integration constants. The relative trajectory described by Equation~\ref{eq:linearTwoBody} is an ellipse in the $x-y$ plane and a closed Lissajous figure in 3D since the phases $\alpha_x$ and $\alpha_z$  are not equal in general. By additionally imposing $\alpha_x = \alpha_z \equiv \alpha_0$ and $\varrho_z =\pm \sqrt{3} \varrho_x$  the relative trajectory becomes a circle around the formation centre which in inclined by $60^\circ$ w.r.t. the orbital plane of the formation centre. Obviously, there are two solutions depending on the sign of $\varrho_z$. These will be called the clockwise $(+)$ and counter-clockwise $(-)$ solutions in this document describing the two possible orientations of the cartwheel triangle\footnote{The former solutions $(+)$ appear to be rotating clockwise as observed from the Sun, while the latter $(-)$ have a counter-clockwise motion. An alternative nomenclature is ``outward ascending'' for the $(+)$ sign and ``outward descending'' for the $(-)$ sign. This nomenclature refers to the direction of the $z$-motion of the spacecraft when it is on its way outward. }.

By placing three spacecraft on this circle with a relative phase of $120^\circ$, an equilateral triangle formation with arm length $d = 2 \sqrt{3} \varrho_x$ is achieved. Using the above equations, the heliocentric orbital elements of the three spacecraft can be derived. They are summarized in Table~\ref{tab:orbElemLinear} .
\begin{table}
\caption{Heliocentric, ecliptic orbital elements for LISA in the linear model}
\label{tab:orbElemLinear}     
\begin{tabular}{llll}
\hline\noalign{\smallskip}
Element & LISA1 & LISA2 & LISA3  \\
\noalign{\smallskip}\hline\noalign{\smallskip}
Semi-major axis        &  $a$ = free & $a$ & $a$ \\
Eccentricity           &  $e=\varrho_x⁄a $ & $e$ & $e$ \\
Inclination            &  $i = \sqrt{3} \varrho_x / a$ & $i$ & $i$ \\
RAAN                   &  $\Omega_1=$ free & $\Omega_2=\Omega_1 + \frac{2\pi}{3}$ & $\Omega_3=\Omega_1 - \frac{2\pi}{3}$ \\
Argument of Perihelion &  $\omega = \begin{cases} \pi/2 \text{(counter-clockwise)} \\ -\pi/2 \text{(clockwise)} \end{cases} $ & $\omega$  & $\omega$ \\
Mean Anomaly           &  $M_1(t=0) =$ free &  $M_2=M_1 - \frac{2\pi}{3}$ & $M_3=M_1 + \frac{2\pi}{3}$ \\
\noalign{\smallskip}\hline
\end{tabular}
\end{table}
In order to fix the relative position of the LISA formation with respect to Earth, the Mean Initial Displacement Angle (MIDA) needs to be chosen as well (cf. section~\ref{subsec:MIDA} for details) as a free parameter to fully define the system. For a model with a circular Earth orbit, the MIDA is identical to the instantaneous initial displacement angle. The difference in the case of an elliptic Earth orbit will be clarified in section~\ref{subsec:MIDA}.
Overall, there are five free continuous and one discrete free parameter in the system:
\begin{enumerate}
	\item arm length, $d$
	\item semi-major axis, $a$
	\item right ascension of the ascending node (RAAN) of LISA1, $\Omega_1$
	\item initial mean anomaly of LISA1, $M_1(t=0)$
	\item argument of perihelion (discrete) $\omega$
	\item MIDA, $\theta_0$
\end{enumerate}
For LISA the semi-major axis would, however, be chosen to be equal to the one of Earth (at least in the linear two-body model) in order to avoid a relative drift.

\subsection{Keplerian two-body model}
\label{subsec:keplerTwoBody}
\begin{figure}
  \includegraphics[width=0.75\textwidth]{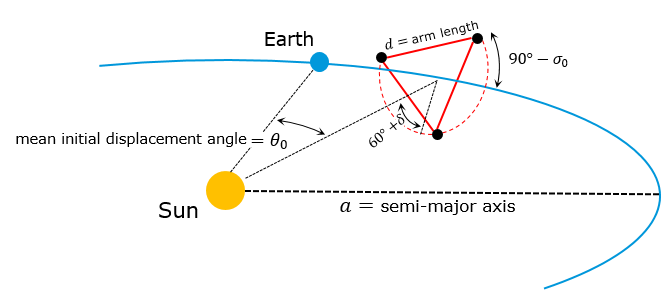}
\caption{Illustration of cartwheel parameters in the Keplerian model.}
\label{fig:cartwheelDef}     
\end{figure}

The model described in section~\ref{subsec:linearTwoBody} can be improved by taking into account the non-linear effects of the Keplerian motion. Still no Earth perturbations are considered here. This is described in reference~\cite{MinimumFlexing}. The cited reference also shows that in this approximation the flexing of LISA's arms can be minimized by deviating slightly from the $60^\circ$ triangle inclination. This deviation is parametrised by the angle, $\delta$. However, it is convenient to instead use the dimensionless parameter $\delta_1$ defined by $\delta = \alpha \delta_1$, with $\alpha=d/2a$. The optimum value is shown to be $\delta_1=5/8$ (this is only true in the Keplerian model). The orbital parameters can be computed from the formulas given in Table~\ref{tab:orbElemKep} where the eccentric anomaly, $E_i$ and the mean anomaly, $M_i$ are related by Kepler's equation:
\begin{equation}
\label{eq:keplerEq}
 E_k + e \sin E_k =	M_k = \pi - \sigma_k(t) , \quad k=1,2,3 \, .
\end{equation}
And the three clocking angles are defined as:
\begin{equation}
	\sigma_k(t) = \sigma_0 + (k-1) \frac{2\pi}{3}  - \sqrt{\frac{\musun}{ a^3 }} t, \quad k=1,2,3 \, ,
\end{equation}
where $\sigma_0$ is called the initial clocking angle.
\begin{table}
\caption{Heliocentric, ecliptic orbital elements for for LISA in the Keplerian model}
\label{tab:orbElemKep}     
\begin{tabular}{llll}
\hline\noalign{\smallskip}
Element & LISA1 & LISA2 & LISA3  \\
\noalign{\smallskip}\hline\noalign{\smallskip}
Semi-major axis        &  $a$ = free & $a$ & $a$ \\
Eccentricity           &  $e = -1 + \sqrt{ 1+ \frac{4}{3}\alpha^2 + \frac{4}{\sqrt{3}} \alpha \cos(\frac{\pi}{3} + \delta ) } $ & $e$ & $e$ \\
Inclination            &  $\tan i  = \frac{2}{\sqrt{3}} \frac{\alpha \sin(\frac{\pi}{3} + \delta ) }{ \left[ 1 + \frac{2}{\sqrt{3}} \alpha \cos(\frac{\pi}{3} + \delta) \right]}$ & $i$ & $i$ \\
RAAN                   &  $\Omega_1 = \sigma_0 - \pi/2$  & $\Omega_2 = \sigma_0 + \pi/6$ & $\Omega_3 = \sigma_0 + 5 \pi/6$ \\
Argument of Perihelion &  $\omega = \begin{cases} \pi/2 \text{ (counter-clockwise)} \\ -\pi/2 \text{ (clockwise)} \end{cases} $ & $\omega$  & $\omega$ \\
Mean Anomaly           &  $M_1(t=0) = \pi - \sigma_0$  &  $M_2=M_1 - \frac{2\pi}{3}$ & $M_3=M_1 + \frac{2\pi}{3}$ \\
\noalign{\smallskip}\hline
\end{tabular}
\end{table}
Overall, there are again five free continuous and one discrete free parameter in the system:
\begin{enumerate}
	\item arm length, $d$
	\item semi-major axis, $a$
	\item inclination parameter, $\delta_1$
	\item initial clocking angle, $\sigma_0 = \sigma_1(t=0)$
	\item argument of perihelion (discrete) $\omega$
	\item MIDA, $\theta_0$
\end{enumerate}
These (except  $\omega$) are illustrated in Figure~\ref{fig:cartwheelDef}.

\begin{figure}
  \includegraphics[width=0.99\textwidth]{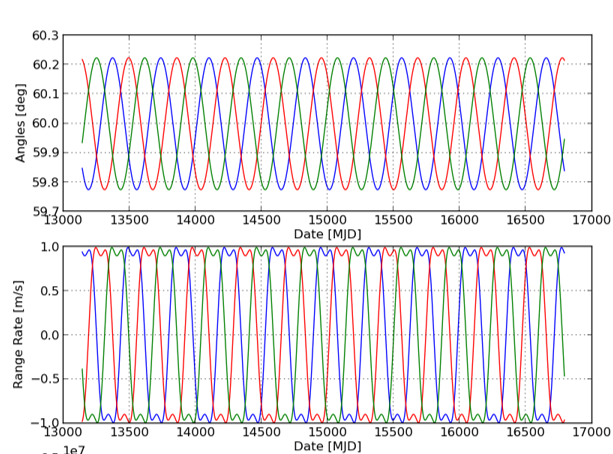}
	\includegraphics[width=0.99\textwidth]{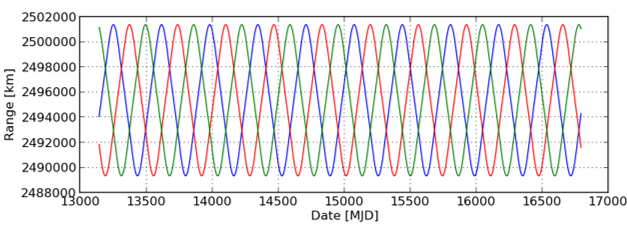}
\caption{Evolution of cartwheel corner angles (top), arm length rate (middle) and arm length (bottom) for the Keplerian carwheel model.}     
\label{fig:breathingKeplerian}
\end{figure}

While the corner angles in the case of the linear model described in section~\ref{subsec:linearTwoBody} are constant at $60^\circ$, the non-linear effects of the Keplerian model introduce a ``breathing'' of the triangle. The resulting evolution of the corner angles, arm length rate and arm length is shown in Figure~\ref{fig:breathingKeplerian} for $a=1 \text{ AU}$, $d=2.5\cdot 10^6 \text{ km}$ and $\delta_1=5/8$.

Although this model is lacking the effect of the Earth gravity perturbation, which is significant for MIDAs close to $20^\circ$, it is still useful as an initial guess for an optimisation using a full numerical model. Moreover, the Keplerian model can be used for the arrival state during the transfer optimisation. This allows for an easy exploration of a large range of orbit options without previous computationally heavy optimisation of the science orbit.

\subsection{Mean Initial Displacement Angle (MIDA)}
\label{subsec:MIDA}
The MIDA, $\theta_0$ defines the location of the centre of the cartwheel with respect to the Earth (see Figure~\ref{fig:cartwheelDef}). A negative MIDA denotes a trailing configuration, a positive MIDA a leading configuration. The larger the MIDA magnitude, the lesser the gravitational perturbations of the cartwheel, but the larger the distance to Earth and thus the impact on the communications subsystem. Conversely, the smaller the MIDA, the lower the $\Delta v$ cost of transfer. A larger MIDA generally also means a higher transfer $\Delta v$, due to the larger difference in semi-major axis of the transfer orbit, if the transfer duration is kept fixed. 

During the science phase the centre of the LISA formation centre moves on a (very close to) circular orbit around the Sun, while the Earth orbit is eccentric. This leads to a natural variation of the instantaneous Earth displacement angle of about $\pm 2^\circ$ over one year. This fact makes the instantaneous displacement angle an inconvenient design parameter when different initial times are compared. In order to compare different launch dates over one year, it is therefore more convenient to use the MIDA, $\theta_0$. The angle averages out the Earth eccentricity by measuring the displacement angle w.r.t. the ``Mean Earth'', a virtual body with the same orbital period as the Earth, but a circular orbit around the Sun (see also ref.~\cite{JoffreICATT}). The orbital elements of the Mean Earth are summarised in Table~\ref{tab:orbElemMeanEarth}.
\begin{table}
\caption{Orbital elements of the Mean Earth from True Earth.}
\label{tab:orbElemMeanEarth}     
\begin{tabular}{ll}
\hline\noalign{\smallskip}
Element &    \\
\noalign{\smallskip}\hline\noalign{\smallskip}
Semi-major axis        &  $a_{\text{mean}} = a_{\text{true}} $ \\
Eccentricity           &  $e_{\text{mean}} = 0.0 $  \\
Inclination            &  $i_{\text{mean}} = i_{\text{true}} $  \\
RAAN                   &  $\Omega_{\text{mean}} = \Omega_{\text{true}} $ \\
Argument of Perihelion &  $\omega_{\text{mean}} = \omega_{\text{true}} $ \\ \\
True Anomaly           &  $\nu_{\text{mean}} = M_{\text{true}} $ \\\\
\noalign{\smallskip}\hline
\end{tabular}
\end{table}

\subsection{Earth perturbations and choice of initial semi-major axis}
\label{subsec:earthPerturb}
The Keplerian model and cartwheel state definition as described in section~\ref{subsec:keplerTwoBody} is suitable for use in transfer optimization and as initial guess for the science orbit optimization. The main perturbation that significantly impacts the LISA science orbit evolution is the Earth's gravity. It is not possible to analytically solve the model when the Earth's gravity is included. However, there are a few useful analytical formulas that can be derived to aid the orbit design. These are described in this section.

\begin{figure}
  \includegraphics[width=0.95\textwidth]{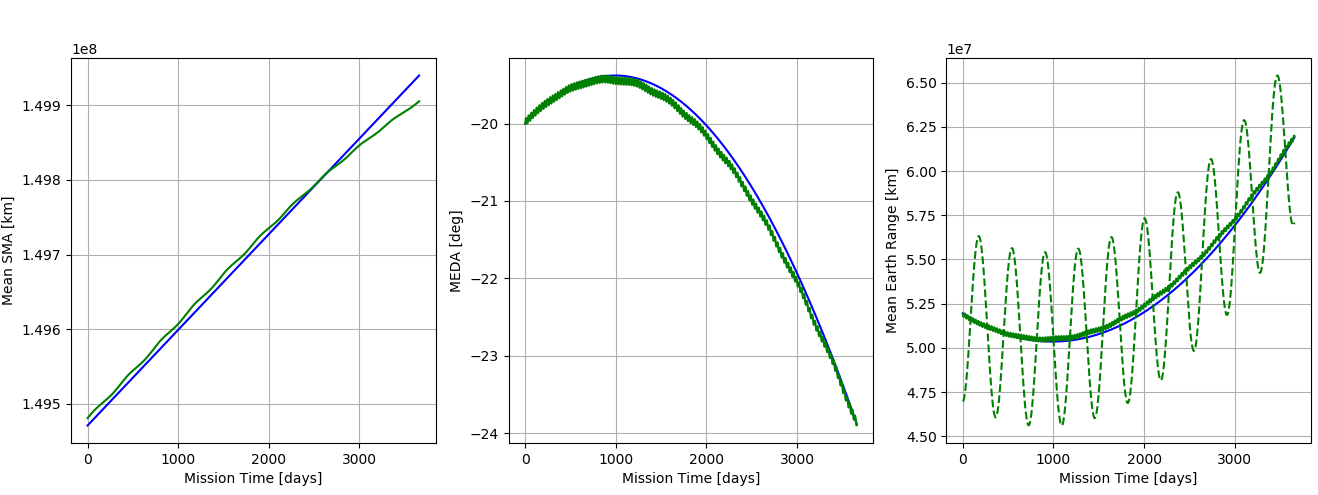}
	\includegraphics[width=0.95\textwidth]{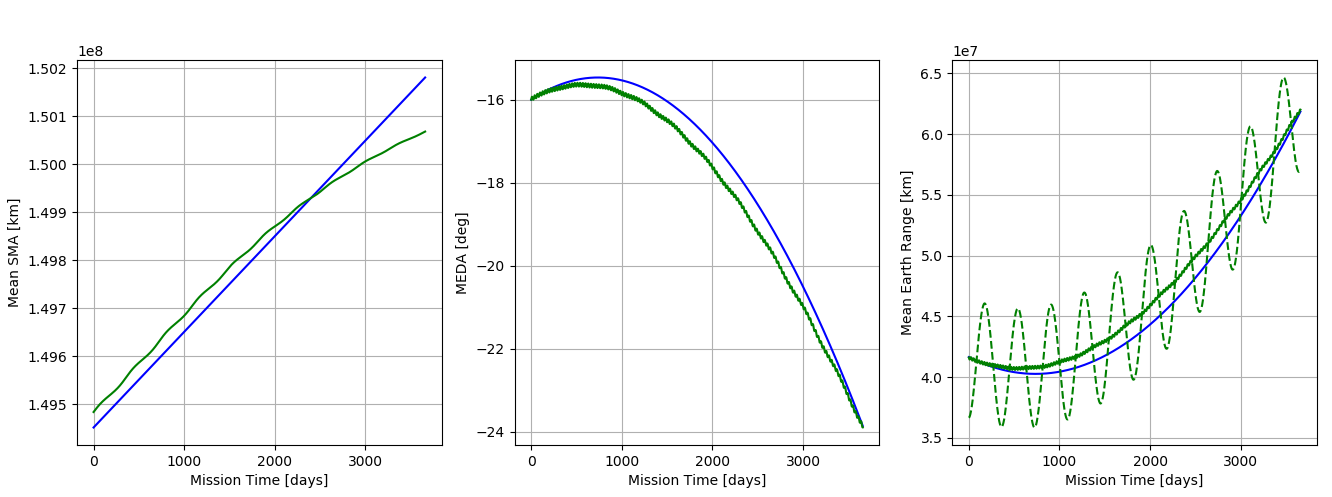}
	\includegraphics[width=0.95\textwidth]{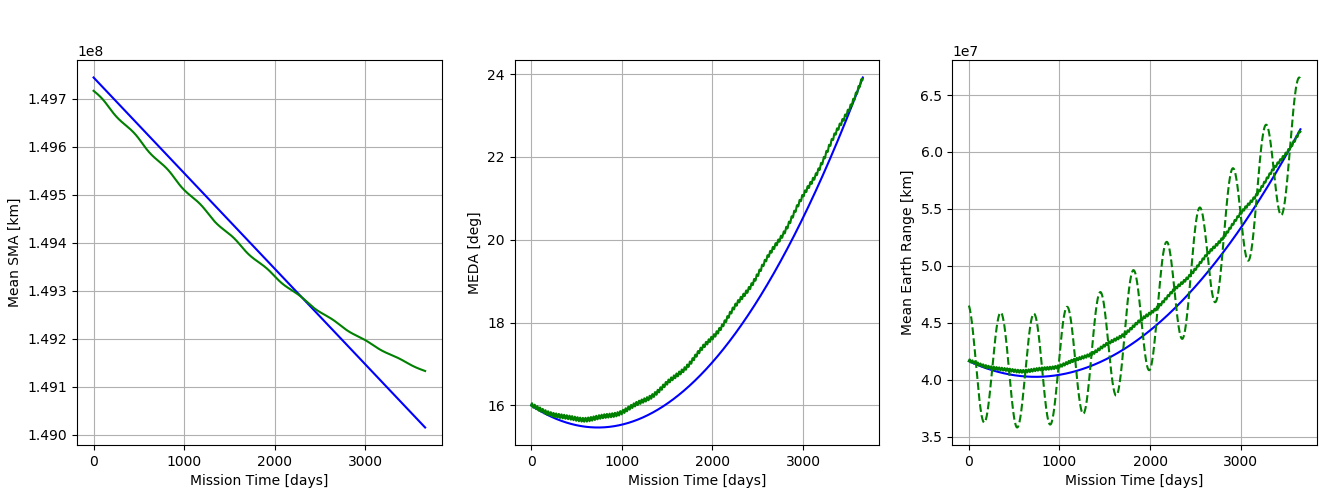}
	\includegraphics[width=0.95\textwidth]{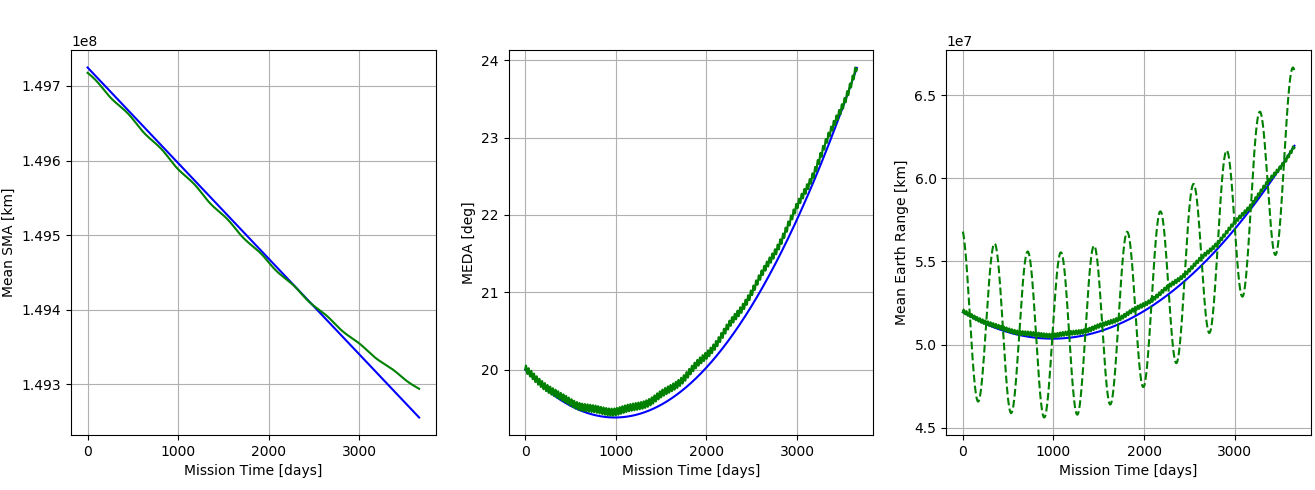}
\caption{Evolution of mean semi-major axis, $a$ (left), mean Earth displacement angle, $\theta$ (middle) and mean Earth range, $r_{\text{Earth}}$ (right), according to approximate equations of section~\ref{subsec:earthPerturb} (blue) for four different MIDA values: $-20^\circ$, $-16^\circ$, $+16^\circ$, $+20^\circ$ (from top to bottom). The green lines show the comparison with a numerical propagation. The dashed green lines show the true Earth range for the plots on the right.}  
\label{fig:analyticalMIDA}
\end{figure}

The main effect of the Earth's gravity (besides perturbing the triangular formation) is a (near) along-track acceleration on the LISA spacecraft which causes a drift in the LISA semi-major axis. The relative difference in semi-major axis to the one of the Earth orbit leads to a drift of the Earth distance. This will be analysed in the following. Assuming circular orbits and neglecting any radial and cross-track contributions, the following formula for the drift rate of the mean semi-major axis can be derived from the Gauss equations:
\begin{equation}
\label{eq:dadt}
	\frac{d a}{d t} = \pm \frac{\muearth}{2 \sin^2(\theta/2) \sqrt{a \musun}} \, .
\end{equation}
The $+$ sign shall be used for the trailing configuration and the -- sign for the leading configuration. Evaluating the equation at $a=1 \text{AU}$ and mean Earth displacement angle $\theta = \theta_0$ gives sufficiently accurate results if $\theta$ is not too far from $20^\circ$. In that case it can also be assumed that the evolution of the mean semi-major axis is strictly linear, i.e. $a(t)=a_0+ \dot{a}t$.
Inserting this evolution of the mean semi-major axis into the (linear) time evolution of the mean anomaly and integrating over time, the time evolution of the mean Earth displacement angle, $\theta$, can be derived:
\begin{equation}
\label{eq:MEDAevol}
	\theta(t) = \theta_0 + \sqrt{\musun} \left( \frac{2}{ \dot{a} \sqrt{a_0} } - \frac{2}{ \dot{a} \sqrt{a_0 + \dot{a} t} } - \frac{t}{\sqrt{\text{AU}}} \right) \, ,
\end{equation}
with $\dot{a}$ given by equation~\ref{eq:dadt}.
The mean Earth distance can be then computed from
\begin{equation}
	r_{\text{Earth}}(t) = 2 a \sin\left( \theta(t)/2 \right) \, .
\end{equation}
The cases where the time evolution of $r_{\text{Earth}}$, has a turning point are the most interesting ones, because they lead to the least difference between the minimum and maximum Earth distance during a given mission time: e.g. for a trailing configuration the initial semi-major axis is chosen slightly smaller than $1$ AU causing an initial decrease of  $r_{\text{Earth}}$. After the turning point is reached at $1$ AU semi-major axis, the Earth distance increases again and reaches its maximum value at the end of the mission. In order to write Equation~\ref{eq:MEDAevol} in a form that illustrates this insight clearly, it can be Taylor expanded in both the semi-major axis change fraction, $\frac{\dot{a} t}{a_0}$, and the deviation of the initial semi-major axis from $1$ AU: $\epsilon \coloneqq 1 - a_0/\text{AU}$
\begin{equation}
	\theta(t) = \theta_0 + \frac{2}{3} \sqrt{\frac{\musun}{\text{AU}^3}} \epsilon t - \frac{3}{4} \sqrt{ \frac{\musun}{\text{AU}^5} } \dot{a} t^2 \, .
\end{equation}
In this form it is clear that the evolution of the mean Earth displacement angle, $\theta$, as a function of time is a parabola. The $\theta$ value reached after the science phase duration, $\Delta t$, is thus basically determined by the choice of the initial semi-major axis, $a_0$. 
This allows computing the initial initial semi-major axis, $a_0$,  from the maximum allowed mean Earth displacement angle, $\theta(\Delta t)$, which occurs at the end of mission:
\begin{equation}
\label{eq:anodFromMIDA}
	a_0 = \text{AU} \left( 1 - \frac{2}{3}\sqrt{\frac{\text{AU}^3}{\musun}} \frac{\theta(\Delta t) - \theta_0}{\Delta t} - \frac{\dot{a} \Delta t}{2 \text{AU}}  \right) \, .
\end{equation}
Since for LISA the constraint on the maximum Earth distance is typically on the true Earth distance and not the mean one, the constant maximum difference between the Earth's mean an true anomaly of almost $1^\circ$ has to be taken into account when computing $\theta(\Delta t)$. With some margin, a value of $1.2^\circ$ has been found to work well for the considered range of MIDAs.
\begin{equation}
\label{eq:finalTheta}
	\theta(\Delta t) = \pm \left( 2 \sin^{-1} \frac{r_{\text{max}}}{2 \text{AU}} - 1.2^\circ\right) \, ,
\end{equation}
where $r_{\text{max}}$ is the maximum allowed true Earth distance.
A graphical representation of the equations in this section for a number of different MIDA values and $r_{\text{max}} = 65 \cdot 10^6 \text{km}$ is shown in Figure~\ref{fig:analyticalMIDA}. The parabolic shape of the mean Earth displacement angle, $\theta$ and mean Earth range evolution is clearly visible and the time where the turning point occurs depends on the choice of the MIDA. The maximum mean Earth range is always reached at the end of mission. The comparison with an exact numerical propagation of the LISA formation centre is also shown. For the numerical propagation the same final $\theta$ value is imposed as for the analytic solution, Equation~\ref{eq:finalTheta}. Note that even for the numeric solution the maximum true Earth distance is not exactly $65 \cdot 10^6 \text{km}$. This is because the constraint is put on the mean Earth distance in this section whereas the actual constraint is on the true Earth distance. This discrepancy can be refined during the actual numeric science orbit optimisation.

Equation~\ref{eq:anodFromMIDA} is particularly useful as it provides an initial guess for the initial semi-major axis of the cartwheel orbit and can also be used in conjunction with the Keplerian model from section~\ref{subsec:keplerTwoBody} for defining the target orbit in the transfer optimization. For the sake of reference, Table~\ref{tab:smaFromMIDA} shows the initial semi-major axis values for a number of MIDA values and the comparison with numerically computed (exact) values. Note that for MIDA values below $\pm 11^\circ$, the constraint on the maximum Earth distance cannot be met any more because initial Earth perturbation is too strong.
\begin{table}
\caption{Optimal values for initial mean semi-major axis according to equation~\ref{eq:anodFromMIDA} and numerically computed value as a function of MIDA, $\theta_0$. The maximum true Earth distance assumed to compute these values is $r_{\text{max}} = 65 \cdot 10^6 \text{km}$.}
\label{tab:smaFromMIDA}       
\begin{tabular}{llll}
\hline\noalign{\smallskip}
MIDA $\theta_0$ [deg] & Optimal initial mean s.m.a.,  & Optimal initial s.m.a., & Difference	\\
                      & $a_0$ (analytical) [km] & $a_0$ (numerical) [km] &  [km]   \\
\noalign{\smallskip}\hline\noalign{\smallskip}
-21.5 & 149460810.7 & 149465317.9 &  -4507.2 \\
-20.0 & 149471018.3 & 149480920.1 &  -9901.8 \\
-18.0 & 149471856.4 & 149491557.8 &  -19701.4 \\
-16.0 & 149451018.2 & 149483335.6 &  -32317.4 \\
-14.0 & 149395265.5 & 149437283.5 &  -42018.0 \\
-12.0 & 149279463.8 & 149237493.4 &  41970.4 \\
+12.0 & 149916277.6 & 149983666.7 &  -67389.1 \\
+14.0 & 149800475.9 & 149765150.1 &  35325.8 \\
+16.0 & 149744723.2 & 149716965.5 &  27757.7 \\
+18.0 & 149723885.0 & 149707659.9 &  16225.1 \\
+20.0 & 149724723.1 & 149717633.0 &  7090.1 \\
+21.5 & 149734930.7 & 149732902.6 &  2028.1 \\
\noalign{\smallskip}\hline
\end{tabular}
\end{table}

The Earth's gravity also has a strong effect on the breathing of the cartwheel and significantly alters the evolution shown for the Keplerian model in Figure~\ref{fig:breathingKeplerian}. This breathing can only be controlled by a careful adjustment of the initial conditions which will be described in the following section.

%%%%%%%%%%%%%%%%%%%%%%%%%%%%%%%%%%%%%%%%%%%%%%%%%%%%%%%%%%%%%%%%%%%
\section{The cartwheel formation - fully numerical model}
\label{sec:cartwheelNumerical}
%%%%%%%%%%%%%%%%%%%%%%%%%%%%%%%%%%%%%%%%%%%%%%%%%%%%%%%%%%%%%%%%%%%
In order to design a realistic cartwheel trajectory, a fully numerical model is required. The dynamical model used here takes into account the point mass gravity of the Sun and all relevant planetary bodies (Mercury, Venus, Earth, Moon, Mars, Jupiter, Saturn). Solar radiation pressure (SRP) and other non-gravitational accelerations are not taken into account as they will be compensated by LISA's  Drag Free and Attitude Control System (DFACS)~\cite{DFACSGath,DFACSKlotz}. This is because the spacecraft trajectories follow the trajectories of the internal test masses which are influenced by gravitational forces only. The effect of the self-gravity acceleration will be discussed later in section~\ref{subsec:selfGravity}.

\subsection{Optimisation for 10 years of science phase}
\label{sec:10yearsOpt}
The six-parameter Keplerian cartwheel model described in section~\ref{subsec:keplerTwoBody} serves as an initial guess for the optimisation. All these parameters, except for the clocking angle, $\sigma_0$, are considered ``design parameters'' here. The assumptions for the values of theses parameters used in this section are shown in Table~\ref{tab:cartwheelParameterAssumptions}. Note that the scientific requirements allow for a range of MIDAs (including Earth-leading configuration) and also for the counter-clockwise option. These options have been analysed internally at ESA, but will not be presented in this paper for brevity.
\begin{table}
\caption{Assumptions for cartwheel design parameters serving as initial guess.}
\label{tab:cartwheelParameterAssumptions}       % Give a unique label
\begin{tabular}{ll}
\hline\noalign{\smallskip}
Parameter & Value \\
\noalign{\smallskip}\hline\noalign{\smallskip}
Arm length, $d$                   & $2.5\cdot 10^6 \text{ km}$ \\
Semi-major axis, $a$              & $149480920.1 \text{ km}$ (from Table~\ref{tab:smaFromMIDA} ) \\
Inclination parameter, $\delta_1$ &  $5/8$\\
Argument of Perihelion, $\omega$   &  $-\pi/2$\\
MIDA, $\theta_0$                  &  $-20^\circ$\\
\noalign{\smallskip}\hline
\end{tabular}
\end{table}
The clocking angle, $\sigma_0$, is considered fixed in the current section as well, but will be used as a free parameter during the transfer optimisation described in section~\ref{subsec:simultaneousOpti}. It is assumed (and verified a-posteriori) that the success of the cartwheel optimisation does not significantly depend on the initial guess of $\sigma_0$. 

The objective of the cartwheel optimisation is to adjust the initial conditions such that the breathing of the operational triangle is controlled within the allowed parameter ranges over 10 years mission time. The optimisation is carried out by allowing all 18 parameters of the initial cartwheel state ($3 \times 6$ Keplerian state parameters) to be adjusted by the optimiser within a user-defined narrow band around the first guess. There are various ways to formulate the problem. Here the approach of using a constant cost function and and adding the stability requirements as constraints has been used. Thus, it is a pure feasibility problem. The assumed constraint definitions are summarised in Table~\ref{tab:cartwheelConstraints}.
\begin{table}
\caption{Constraints in the cartwheel optimisation problem.}
\label{tab:cartwheelConstraints}       % Give a unique label
\begin{tabular}{ll}
\hline\noalign{\smallskip}
Constraint     & Allowed range \\
\noalign{\smallskip}\hline\noalign{\smallskip}
Minimum corner angle           & $> 59.0^\circ$ \\
Maximum corner angle           & $< 61.0^\circ$ \\
Minimum arm length             & $> 2.49 \cdot 10^6 \text{ km}$ \\
Maximum arm length             & $< 2.51 \cdot 10^6 \text{ km}$ \\
Minimum arm length rate        & $> -10 \text{ m/s}$ \\
Maximum arm length rate        & $< 10 \text{ m/s}$ \\
MIDA                           & $ 19.9^\circ < \theta_0 < 20.1^\circ$ \\
Maximum Earth distance of cartwheel centre   &  $< 65 \cdot 10^6 \text{ km}$ \\
\noalign{\smallskip}\hline
\end{tabular}
\end{table}

The propagation of the initial state was carried out using a Runge-Kutta (8)7 dense stepper~\cite{denseRungeKutta1,denseRungeKutta2}. It also propagates the state transition matrix which is used together with an in-house automatic differentiation software to obtain analytical partials for the optimisation. For the actual optimisation SNOPT~\cite{snopt} is called via the optimisation framework PyGMO~\cite{pygmo} from Python. The minimum and maximum values over the mission time used for the constraints evaluation (Table~\ref{tab:cartwheelConstraints}) are obtained from sampling the trajectory with a step size of 10 days.

\begin{figure}
  \includegraphics[width=0.99\textwidth]{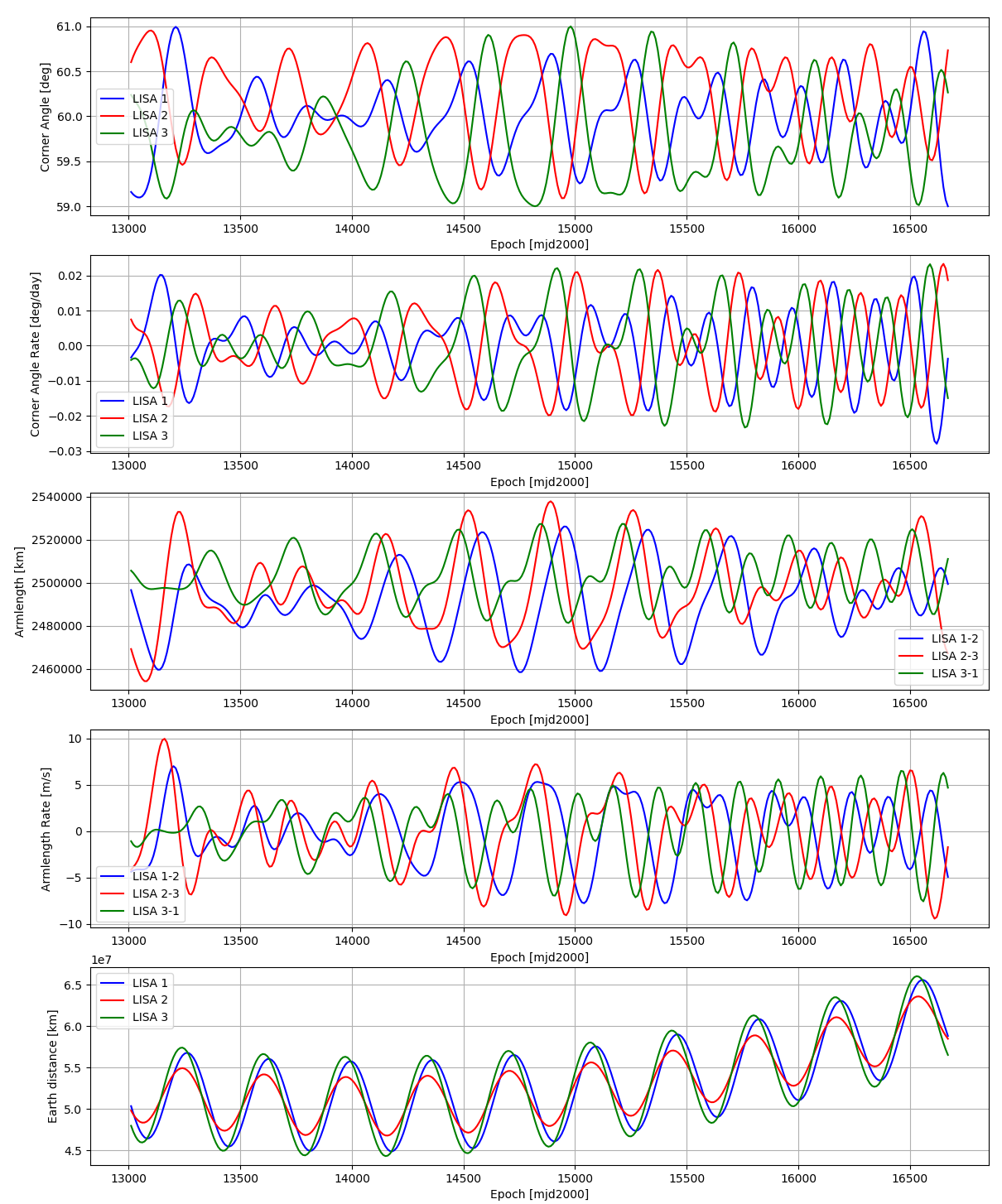}
\caption{Evolution of the cartwheel formation over 10 years in a fully numerical model.}  
\label{fig:cartwheelGeom}
\end{figure}

The resulting evolution of the cartwheel geometry is shown in Figure~\ref{fig:cartwheelGeom}. It can be seen that the corner angles and arm length rates can be constrained within the required windows over the considered mission time of 10 years. The evolution of the Earth range shows a characteristic profile where the maximum distance of $65 \cdot 10^6 \text{ km}$ to the formation centre is reached at the end of mission. It is a result of the initial semi-major axis choice (cf. section~\ref{subsec:earthPerturb}) which leads to an initial drift towards Earth followed by a turning point. Note that the initial semi-major axis determines the Earth distance profile and thus the overall impact of the Earth perturbation on the formation stability. Relaxing the maximum allowed Earth distance decreases the Earth perturbation and thus allows choosing more narrow windows for the corner angles and arm length rates. E.g. by increasing the maximum Earth distance constraint from $65 \cdot 10^6 \text{ km}$ to $75 \cdot 10^6 \text{ km}$ it is possible to reduce the corner angle variations from $\pm 1.0^\circ$ to	$\pm 0.75^\circ$.

\begin{figure}
  \includegraphics[width=0.99\textwidth]{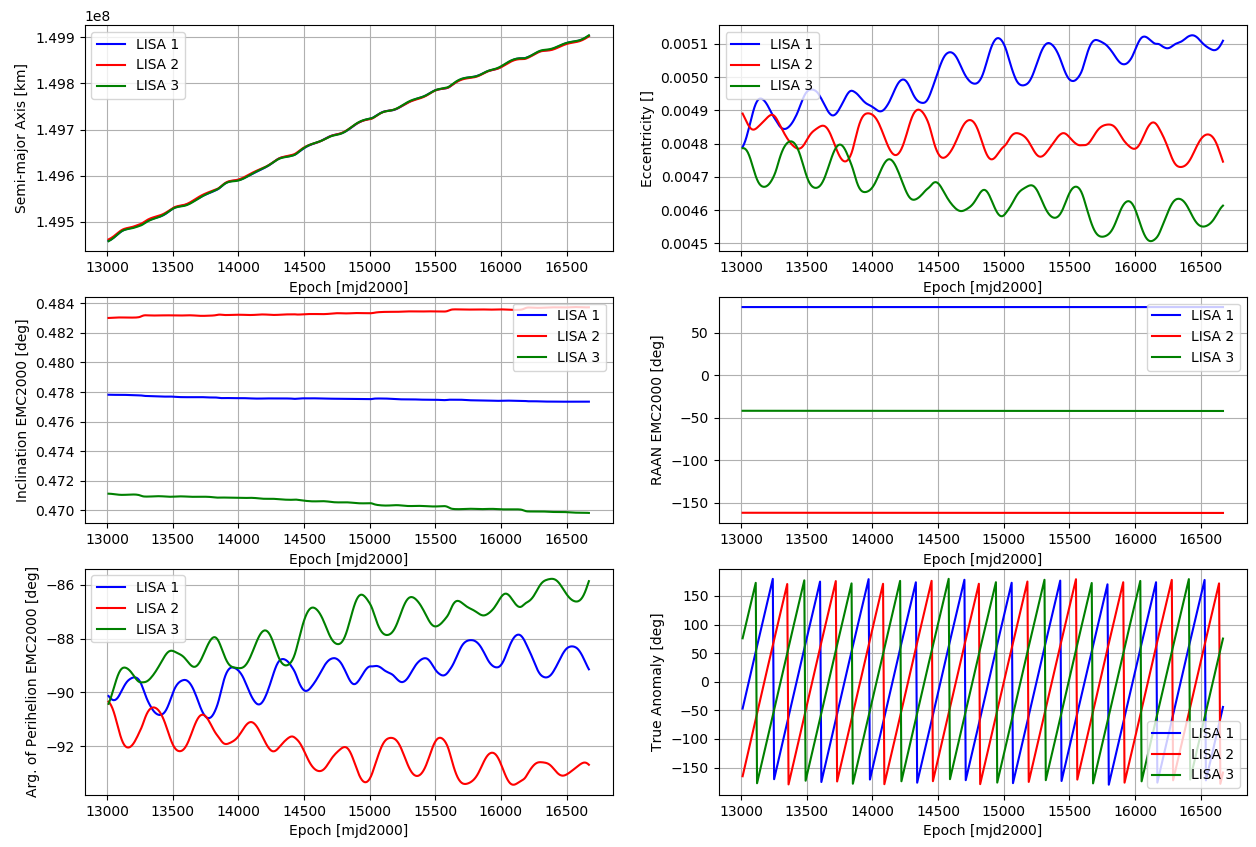}
\caption{Evolution of the cartwheel orbital elements in the ecliptic frame over 10 years in a fully numerical model.}  
\label{fig:cartwheelElem}
\end{figure}

Figure~\ref{fig:cartwheelElem} shows the evolution of the Keplerian elements during the 10 years science phase. The drift of the semi-major axes due to the Earth's influence is apparent. Moreover, the initial values of the other elements are seen to be close to the initial guess determined by the Keplerian model (cf. section~\ref{subsec:keplerTwoBody}), but drifting away due to the third body perturbations.

The initial Keplerian parameters of the cartwheel orbit shown in Figures~\ref{fig:cartwheelGeom} and~\ref{fig:cartwheelElem} are given in Table~\ref{tab:cartwheelOptParam} for reference.

\begin{table}
\caption{Initial osculating keplerian state in EME2000 for the cartwheel orbit obtained in the fully numerical model. The reference epoch is 2035-08-15T12:00:00 TDB.}
\label{tab:cartwheelOptParam}       % Give a unique label
\begin{tabular}{llll}
\hline\noalign{\smallskip}
Spacecraft & LISA1 & LISA2 & LISA3  \\
\noalign{\smallskip}\hline\noalign{\smallskip}
Semi-major axis [km]         & 149461821.067 & 149458004.472 & 149458683.642 \\
Eccentricity [-]             & 0.0048903     & 0.0047861     & 0.0047896 \\
Inclination [deg]            & 22.9818       & 23.7923       & 23.5280   \\
RAAN [deg]                   & -0.3933       & -0.7788       & 1.1782    \\
Argument of perihelion [deg] & 108.5557      & 131.5359      & -11.3762  \\
True anomaly [deg]           & -165.1558     & 75.9475       & -46.9373   \\
\noalign{\smallskip}\hline
\end{tabular}
\end{table}

\begin{figure}
  \includegraphics[width=0.7\textwidth]{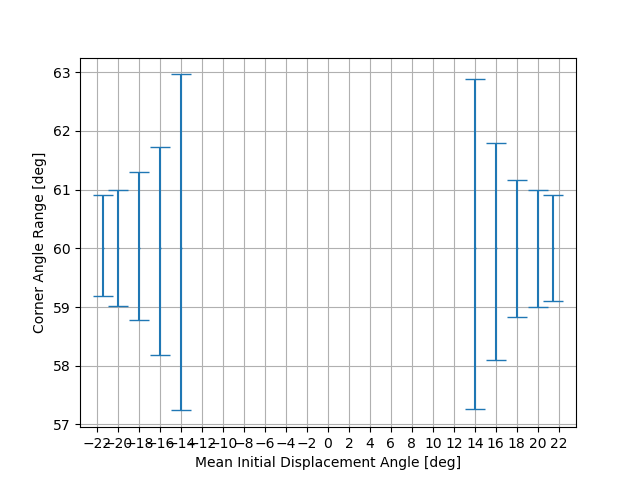}
\caption{Achievable corner angle variations for cartwheel orbits with different MIDA values. The mean arm length and maximum Earth distance are kept at $2.5 \cdot 10^6 \text{ km}$ and $65 \cdot 10^6 \text{ km}$, respectively.}  
\label{fig:cornerAngleRanges}
\end{figure}

Naturally, the chosen MIDA also has a strong impact on the obtainable corner angle variations, because it determines the initial strength of the Earth's gravitational perturbation. A parametric analysis on different MIDA values is shown in Figure~\ref{fig:cornerAngleRanges}. For MIDA values below $\pm 14^\circ$ it was very difficult to achieve convergence at all therefore the solutions don't appear in the plot.

\subsection{Spacecraft self-gravity}
\label{subsec:selfGravity}
The analysis in the previous section neglected one important dynamical effect, namely the spacecraft self-gravity. This is an effective acceleration component acting on the spacecraft resulting from the DFACS following the motion of the two internal test masses which are not located precisely in the centre of mass of the spacecraft. To understand this effect, first consider a single test mass which is offset from the spacecraft centre of mass. The mass of the spacecraft will exert a small gravitational force on the test mass and will cause it to move towards the spacecraft centre of mass. This motion will be detected by the internal interferometers (between the test mass and the optical bench) and will command the DFACS to accelerate the spacecraft into the opposite direction such that the test mass effectively doesn't move with respect to the spacecraft.

Now LISA does not have one, but two test masses per spacecraft and the situation becomes slightly more complicated: since the two test masses are at different locations with respect to the spacecraft centre of mass, there will also be a relative gravitational acceleration between them (in addition to the common-mode acceleration described above). It is thus not possible for the DFACS to follow both test masses at the same time if both are free-falling. The solution is to only have the test masses free-falling along one space dimension (along the laser arm). The self-gravity acceleration on the test masses perpendicular to their laser arm directions is compensated using electrostatic actuators. The reaction forces of these electrostatic actuations on the spacecraft also need to be compensated by the DFACS. This adds another component to the net spacecraft acceleration and needs to be taken into account in the orbit propagation. To summarize, there are two self-gravity acceleration components: a common-mode component, which is also present with only one test mass, and a differential component which results from a relative acceleration between the test masses.

\begin{figure}
  \includegraphics[width=0.5\textwidth]{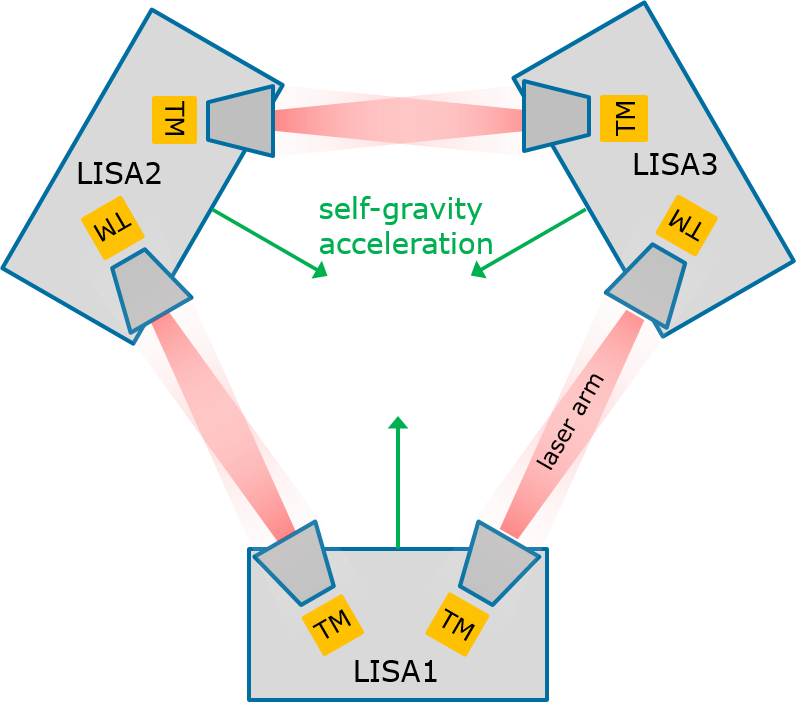}
\caption{Schematic representation of the LISA formation and direction of the self-gravity acceleration. In reality the self-gravity acceleration can be directed in any of the three spacecraft axes.}    
\label{fig:self-gravity}
\end{figure}

The total self-gravity acceleration depends on the spacecraft configuration which is not known at this point of the project. Theoretically, it can be directed along any of the three spacecraft axes. For simplicity, an analysis has been conducted with a self-gravity acceleration only towards the cartwheel centre (cf. Figure~\ref{fig:self-gravity}). A more detailed Monte-Carlo analysis allowing any direction for the acceleration will be presented in section~\ref{subsec:stabilitySelfGravity}.

\begin{figure}
  \includegraphics[width=0.99\textwidth]{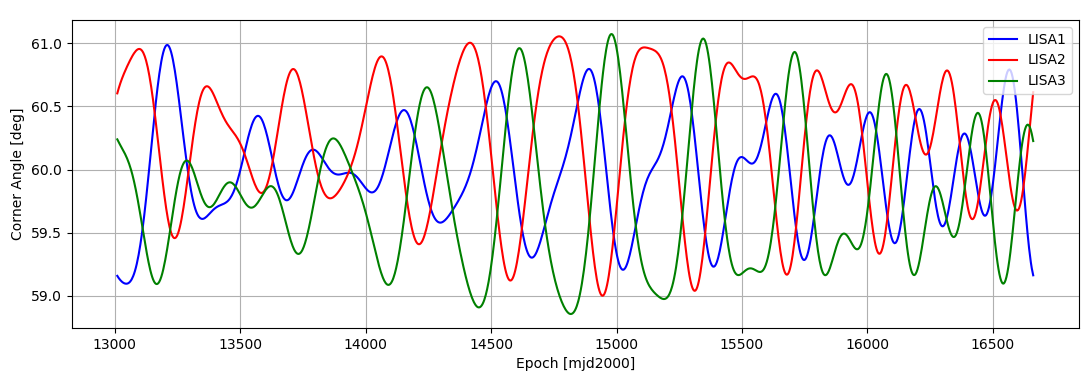}
	\includegraphics[width=0.99\textwidth]{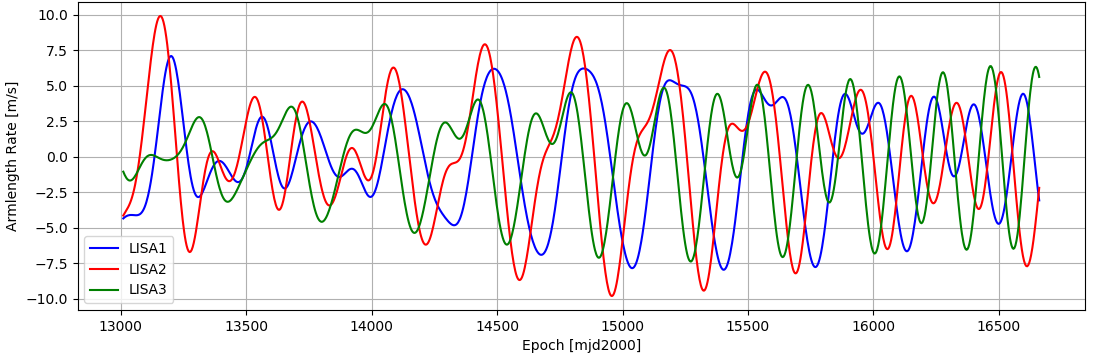}
\caption{Corner angles evolution (top) and arm length rate evolution (bottom) for added self-gravity acceleration linearly varying from $-2 \text{ nm/s}^2$ to $+2 \text{ nm/s}^2$ in the direction of the cartwheel centre.}    
\label{fig:self-gravityEvol2}
\end{figure}

\begin{figure}
  \includegraphics[width=0.99\textwidth]{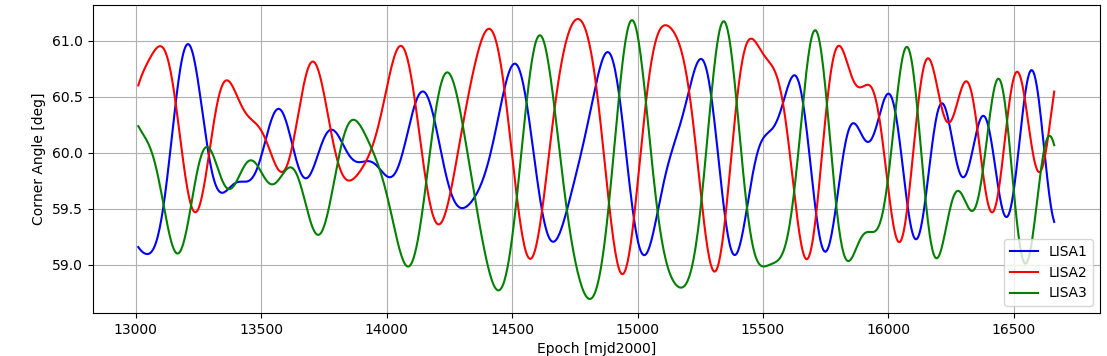}
	\includegraphics[width=0.99\textwidth]{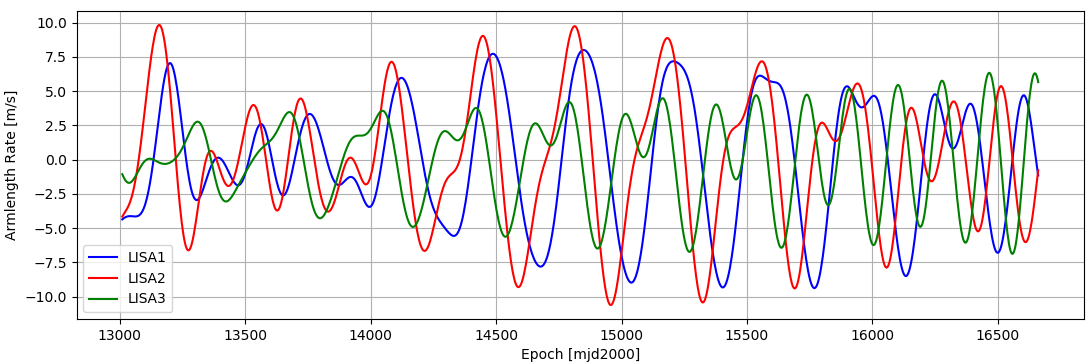}
\caption{Corner angles evolution (top) and arm length rate evolution (bottom) for added self-gravity acceleration linearly varying from $-4 \text{ nm/s}^2$ to $+4 \text{ nm/s}^2$ in the direction of the cartwheel centre.}    
\label{fig:self-gravityEvol4}
\end{figure}

Moreover, due to fuel depletion the magnitude of the acceleration is not constant during the mission time. A detailed model of the acceleration history is yet to be developed and is not subject of the present paper. For simplicity, the current assessment considers acceleration levels changing from $-2 \text{ nm/s}^2$ (start of mission) to $+2 \text{ nm/s}^2$ (end of mission) where a positive value means an acceleration towards the cartwheel centre. Also the range of $-4 \text{ nm/s}^2$ (start of mission) to $+4 \text{ nm/s}^2$ (end of mission) has been looked at for comparison. The state from Table~\ref{tab:cartwheelOptParam} has been propagated taking into account these accelerations and a comparison of the corner angles and arm lengths rate evolution is shown in Figures~\ref{fig:self-gravityEvol2} and~\ref{fig:self-gravityEvol4}, respectively.
 
The first case shows a mild violation of the corner angle constraints of $ \pm 1.0^\circ$ allowed variation. In the second case the violation becomes more severe. Note that no further optimisation has been conducted here. Once a more detailed model of the self-gravity acceleration history is available, it can be implemented as part of the optimisation procedure described in section~\ref{sec:10yearsOpt}. It is expected that accelerations of the levels considered here can be compensated for by a further adjustment of the initial cartwheel state and a stable formation within the requirements can be achieved. 

It is considered more problematic if the self-gravity acceleration has a significant unknown contribution (along all three axes). Such a contribution can lead to a violation of the formation constraints, because it cannot be taken into account operationally in the final manoeuvre optimisation at the time of cartwheel orbit insertion. Its effect will be assessed in section~\ref{subsec:stabilitySelfGravity}.

\subsection{Station keeping}
\label{subsec:stationKeeping}
Nominally, no station keeping manoeuvres are foreseen during the 10 years science phase. The operational orbit and insertion conditions are required to be designed such that the requirements on the corner angle variations and arm length rates are fulfilled. Whether this is possible or not mainly depends on the chosen MIDA value as was shown in Figure~\ref{fig:cornerAngleRanges}. The plot shows that if the corner angles are required to stay within a $\pm 1.0^\circ$ window, the MIDA cannot be below $\pm 20^\circ$. 
However, going to lower MIDA values would have the advantage of a reduced transfer \dv~(cf. section~\ref{subsec:simultaneousOpti}). Therefore, if MIDA values lower than $\pm 20^\circ$ shall be employed, station keeping is required. Note again that these conclusions hold for a maximum allowed Earth distance of $65\times 10^6$ km. For larger values of that constraint it should be possible to reduce the MIDA below $\pm 20^\circ$ without requiring station keeping.

\begin{figure}
  \includegraphics[width=0.5\textwidth]{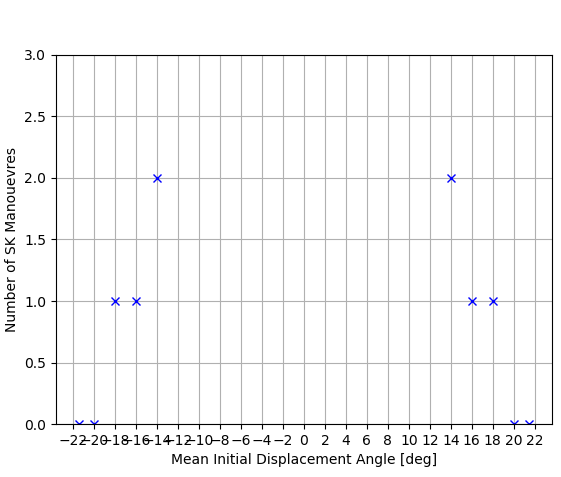}
	\includegraphics[width=0.5\textwidth]{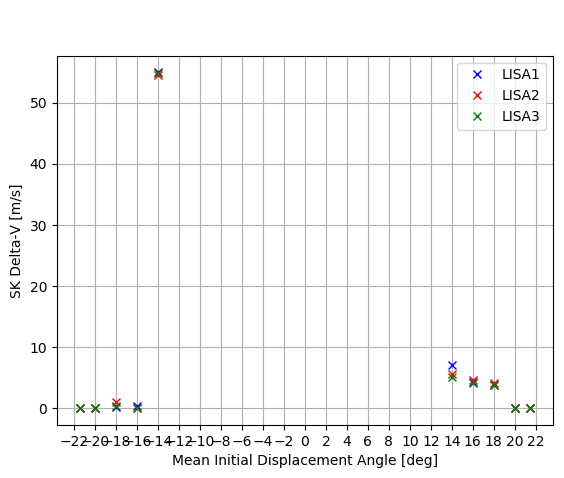}
\caption{Number of required station keeping maneuvers (left) and total station keeping \dv~(right) as a function of the MIDA.}    
\label{fig:stationKeeping}
\end{figure}

Different station keeping strategies are possible. Manoeuvres imply an interruption of the science operations, and require a re-acquisition of the formation. Therefore, it is desirable to minimize the number of manoeuvres and also perform them simultaneously with all 3 spacecraft, if possible. Gaps in the science operations of longer than one week per year are not permitted. The simplest station keeping strategy that complies with these requirements is to place simultaneous manoeuvres after regular time intervals. The number of manoeuvres required to stabilize the orbit for 10 years as a function of the MIDA value is shown in Figure~\ref{fig:stationKeeping} (left). 

The right panel in that figure shows the total required station keeping \dv. The corner angles were required to stay in a window of $\pm 1.0^\circ$. No additional objectives like change of the Earth range drift rate were imposed. The manoeuvres were equally spaced in time during the 10 years science duration, i.e. for the case of 2 manoeuvres, they take place 3.33 and 6.66 years after science orbit insertion.
Note that there is an outlier for the $\theta_0=-14^\circ$ case. This is most likely due to the simple manoeuvre strategy where the manoeuvre times are not optimized. For all the other cases a station keeping budget of $10 \text{ m/s}$ per spacecraft is sufficient.
In case, the \dv~budget requires going below a MIDA of $\pm 20^\circ$, it is recommended to not go lower than $\pm 16^\circ$, which still is possible with a single station keeping manoeuvre after 5 years. Note however, this does not yet include perturbations that come from science orbit insertion accuracy. That contribution has been analysed in more detail in section~\ref{subsec:stability} and~\ref{subsec:stabilitySelfGravity}.

%%%%%%%%%%%%%%%%%%%%%%%%%%%%%%%%%%%%%%%%%%%%%%%%%%%%%%%%%%%%%%%%%%%
\section{Solar electric propulsion transfer}
\label{sec:SEPtransfer}
%%%%%%%%%%%%%%%%%%%%%%%%%%%%%%%%%%%%%%%%%%%%%%%%%%%%%%%%%%%%%%%%%%%
The current section is going to describe the transfer from launch to cartwheel orbit insertion using Solar electric propulsion (SEP). Since the transfer \dv~depends on the cartwheel clocking angle at the time of insertion (cf. section~\ref{subsec:simultaneousOpti}), both transfer and science phase cannot be strictly separated. Section~\ref{subsec:simultaneousOpti} will describe the method that was used to take the interconnection into account.

\subsection{Direct escape launch}
\label{subsec:launch}
A joint launch of all three spacecraft from Kourou into direct escape is the current baseline for ESA. Currently, an all-year launch window is targeted. In order to limit the number of required launcher programs to one, a single separation state (defined in the Earth-fixed) frame has been assumed. Since at the time of this analysis the trajectory analysis from the launch service provider is still under consolidation, a GTO-like escape has been assumed. For simplicity, the assumed orbital parameters at virtual perigee\footnote{The state at virtual perigee is obtained by a Keplerian backwards propagation of the launcher separation state to perigee.} are defined in the inertial EME2000 reference frame and are summarised in Table~\ref{tab:EscapeElements}. The RAAN is a free optimisation parameter and can be fixed by the appropriate choice of the lift-off hour.

\begin{table}
\caption{Assumptions on the escape trajectory in EME2000 reference frame.}
\label{tab:EscapeElements}       % Give a unique label
\begin{tabular}{lll}
\hline\noalign{\smallskip}
Parameter & Value   \\
\noalign{\smallskip}\hline\noalign{\smallskip}
Perigee radius & $6628.14 \text{ km}$  \\
Infinite velocity & $300 \text{ m/s}$  \\
Inclination & $7.0^\circ$  \\
RAAN & free  \\
Argument of perigee & $180.0^\circ$  \\
True anomaly & $0.0^\circ$  \\
\noalign{\smallskip}\hline
\end{tabular}
\end{table}

After separation from the launcher upper stage, the three spacecraft follow independent trajectories, which initially will still be closely grouped.

The SEP is assumed to be used at the earliest 30 days after launch in order to allow for sufficient commissioning time. The actual start time of the first manoeuvre follows from the parametric optimization process of each orbit and can take place much later after launch than 30 days.

\subsection{Simultaneous optimisation of transfer and science orbit}
\label{subsec:simultaneousOpti}
The simplest strategy would be to completely decouple the transfer optimisation from the cartwheel orbit optimisation and assume a transfer to a fixed cartwheel orbit. However, the transfer \dv~is expected to depend on the chosen cartwheel orbit, in particular the initial clocking angle, $\sigma_0$. This dependency for the February 2034 launch will be discussed in more detail in section~\ref{subsec:FebruaryLaunch} and shown in Figure~\ref{fig:clockingAngleSens}. Since a end-to-end optimisation in the full dynamical model may be difficult to implement in a robust way, the following three-stage strategy has been employed:
\begin{enumerate}
	\item Assume a Keplerian cartwheel orbit and optimise the transfer with the clocking angle, $\sigma_0$ as a free parameter.
	\item Optimise the cartwheel orbit with full dynamics for the obtained $\sigma_0$ value.
	\item Re-optimise the transfer assuming the obtained cartwheel orbit as fixed.
\end{enumerate}
The change in transfer \dv~between state 1 and stage 3 is expected to be small which will be confirmed later on. Stage 2 has already been discussed in section~\ref{sec:cartwheelNumerical}. The following will focus on stage 1.

The individual manoeuvre time-line is optimized for each spacecraft such that its target orbit is met with minimal propellant expenditure. All three spacecraft need to be optimised together since they share the launch vehicle and therefore their (free) launch RAAN needs to be the same. A transfer duration of less than 540 days is targeted. This constraint implies that transfers with more than one revolution around the Sun cannot be used. During SEP operations, the Sun aspect angle (SAA, defined as the angle between the direction of the current thrust acceleration vector and the direction from the spacecraft towards the Sun) must remain within a range of $90^\circ \pm 40^\circ$. This is required in order to guarantee sufficient illumination of the solar arrays, knowing that the solar array normal is perpendicular to the thruster direction.

The starting assumptions for the SEP transfer to the cartwheel orbits are listed in Table~\ref{tab:sepAssumptions}. In order to account for outages during thrust arcs due to ground communications and contingencies, a duty cycle of $90\%$ has been assumed. This has been modelled simply as a factor on the nominal available thrust level. Hereafter always the effective thrust level (i.e. nominal thrust times duty cycle) will be used to label cases. 

The assumed wet mass is compatible with an Ariane 64 launcher performance of $7000 \text{ kg}$ into direct escape with an infinite velocity of $300 \text{m/s}$. The maximum \dv~of the three spacecraft is used as an objective function during the optimisation process. This number is relevant for the tank sizing since all three spacecraft are going to be built identically. 

The transfer has been transcribed to a multiple shooting problem allowing up to four thrust arcs per spacecraft. The thrust direction of each thrust arc is parametrised by two angles and their rates which are free parameters. The thrust magnitude has been assumed constant. WORHP~\cite{worhp} has been used from PyGMO~\cite{pygmo} as an optimisation algorithm. Initial guesses for the optimisation have been generated based on simple Lambert arc solutions with impulsive manoeuvres. Starting from these, a series of optimisation runs have been conducted for each trajectory, successively decreasing the thrust magnitude to the target value of $81$ mN. Moreover, the transfer optimisation has first been done for each spacecraft individually keeping the launch RAAN and the clocking angle fixed to the initial guess value. In a second step the launch RAAN and the clocking angle have been freed and all three spacecraft have been re-optimised together to arrive at a globally optimal transfer solution.

\begin{table}
\caption{Assumptions for SEP transfers}
\label{tab:sepAssumptions}      
\begin{tabular}{ll}
\hline\noalign{\smallskip}
Quantity & Value \\
\noalign{\smallskip}\hline\noalign{\smallskip}
Wet mass 	& $2200 \text{ kg}$ \\
Nominal thrust level 	& $90\text{ mN}$ \\
Duty cycle	& $90\%$  \\
Specific impulse  &	$1660\text{ s}$ \\
Minimum Sun aspect angle (SAA) during thrust &	$50^\circ$ \\
Maximum SAA during thrust	& $130^\circ$ \\
Minimum time until first thrust arc	& $30\text{ days}$ \\
Maximum transfer duration	& $540 \text{ days}$ \\
\noalign{\smallskip}\hline
\end{tabular}
\end{table}

\subsection{February 2034 launch to MIDA$=-20^\circ$ cartwheel}
\label{subsec:FebruaryLaunch}
\begin{figure}
  \includegraphics[width=0.5\textwidth]{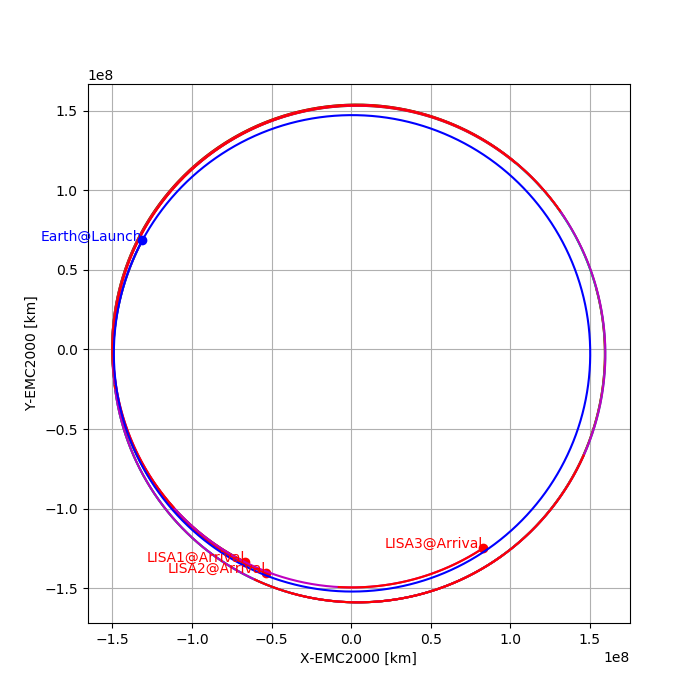}
	\includegraphics[width=0.5\textwidth]{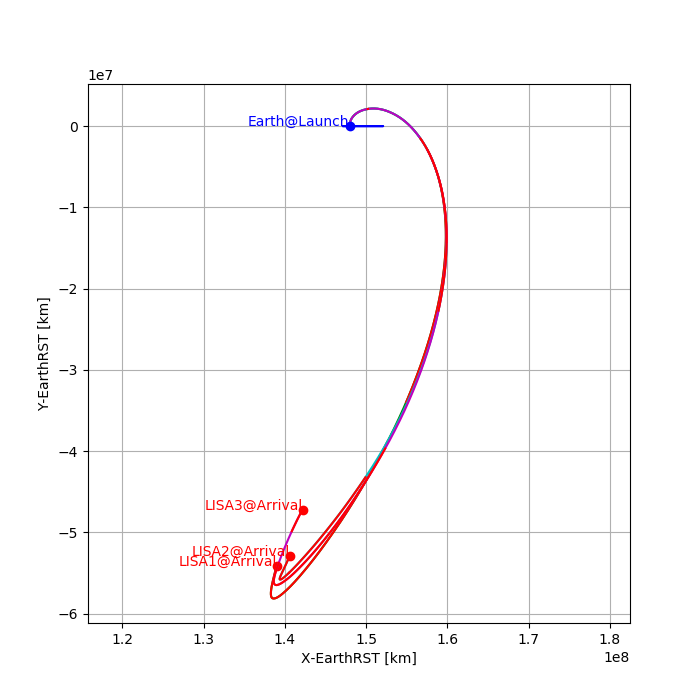}
\caption{Trajectory plots for the February launch transfer in the Ecliptic Frame (left) and the Earth local orbital frame (right). }    
\label{fig:transferTrajectory}
\end{figure}

\begin{figure}
  \includegraphics[width=0.99\textwidth]{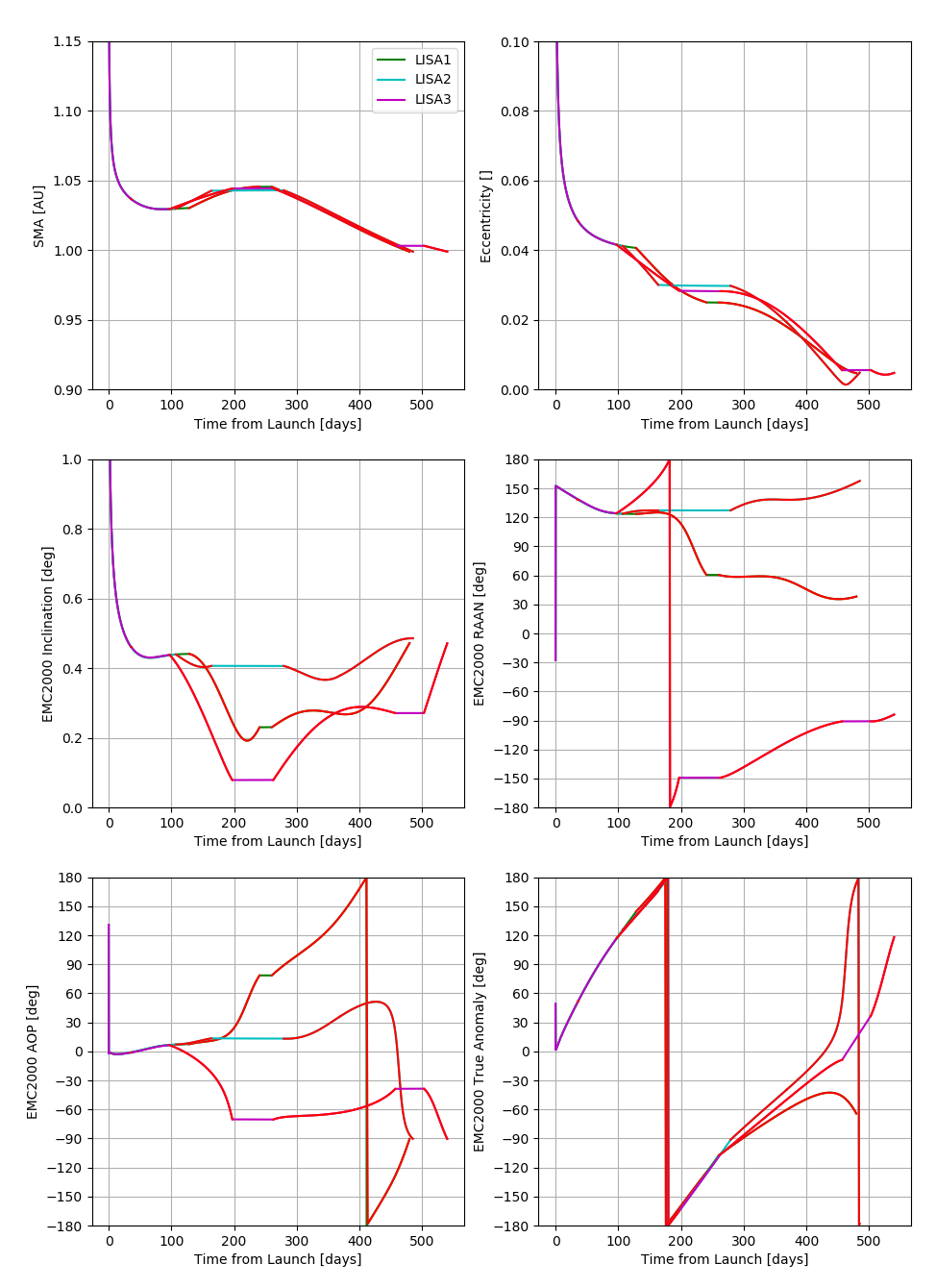}
\caption{Evolution of the osculating orbital elements for the February launch option. }    
\label{fig:transferElem}
\end{figure}

\begin{figure}
  \includegraphics[width=0.99\textwidth]{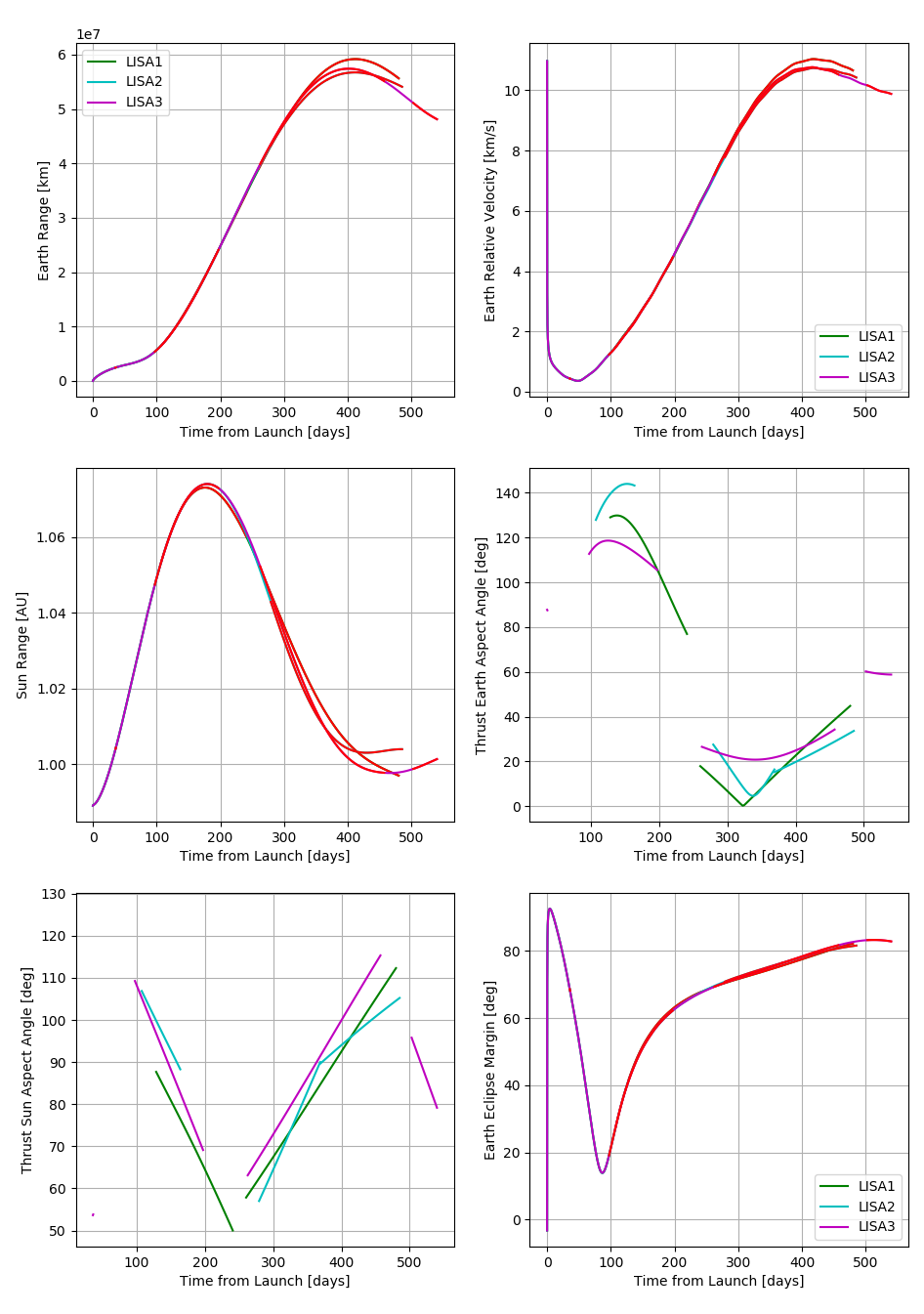}
\caption{Evolution of the Earth, Sun and thrust direction geometry for the February launch option.}    
\label{fig:transferGeom}
\end{figure}

To illustrate the SEP transfer in detail, the launch epoch of 2034-02-21 12:00 is chosen as an example. In a later analysis of 12 representative launch epochs per year the February launch turns out to be the sizing case for 2034 in terms of~\dv. The targeted operational orbit is one with a MIDA of $-20^\circ$ with parameter assumptions as shown in Table~\ref{tab:cartwheelParameterAssumptions}. Up to four thrust arcs are allowed for the 1-revolution transfer. The SEP transfer trajectory for an effective thrust of 81 mN is shown in Figure~\ref{fig:transferTrajectory} (left) in the Ecliptic Reference Frame of epoch 2000. The Earth orbit is displayed in blue, thrust arcs are red and the arrival points are indicated by red circles. The right plot in the Earth local orbital frame illustrates the path that LISA takes w.r.t. Earth. The Earth local orbital frame is defined as follows:
\begin{itemize}
	\item X-axis along the Sun-Earth line
	\item Z-axis along the Earth angular momentum vector
	\item Y-axis completing the right-handed system
\end{itemize}
For the Earth-trailing option the transfer will always first increase the semi-major axis in order to achieve the correct phasing. This implies that the spacecraft can pass through eclipses by the Earth in the early transfer phase especially for launch dates close to the equinoxes. The transfer in February in eclipse-free.
The evolution of the orbital elements for the March launch are shown in the Figure~\ref{fig:transferElem}. The arrival state fulfils the conditions described in section~\ref{subsec:keplerTwoBody}.
The evolution of the geometry w.r.t. to the Sun and Earth is shown in Figure~\ref{fig:transferGeom}. The Earth eclipse margin is an angle quantifying the eclipse condition. If this angle becomes negative, the spacecraft enters eclipse.

\begin{figure}
  \includegraphics[width=0.99\textwidth]{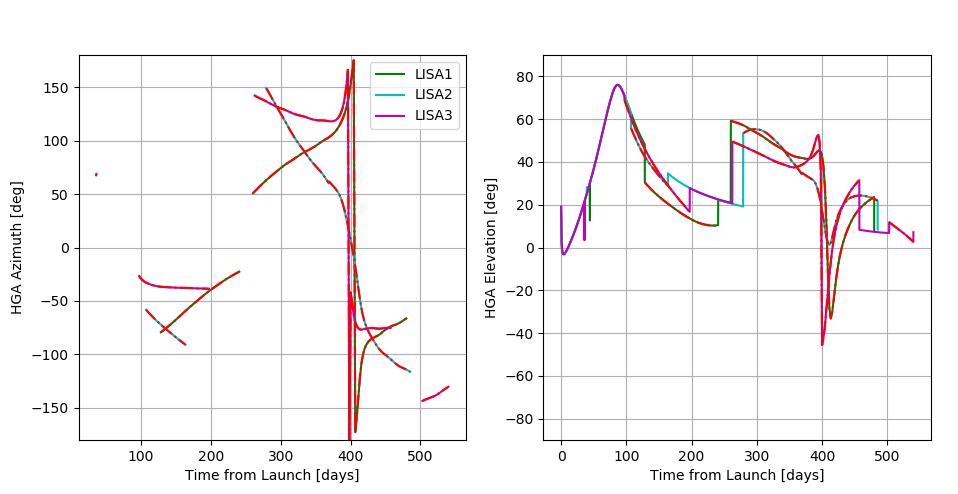}
\caption{HGA pointing evolution for the February launch option.}    
\label{fig:HGApointing}
\end{figure}

An important aspect for communications is the pointing direction of the High-Gain Antenna (HGA). For this analysis it is assumed that the antenna is mounted on the opposite side from the solar array which is facing the spacecraft $+$Z axis. Therefore, if the HGA elevation becomes too large, the field of view is obstructed by the spacecraft. A maximum allowed elevation of $+25^\circ$ is assumed for the time being. The attitude of the spacecraft during transfer is assumed as follows:
\begin{itemize}
	\item During thrust arcs the acceleration vector is always along the spacecraft $+$Y axis.
	\item The Sun incident angle on the spacecraft $+$Z axis (solar array) is maximized. For coast arcs this means that the Sun direction is always along the $+$Z axis.
	\item The spacecraft X axis completes the right-handed system.
\end{itemize}
During coast arcs this attitude is not completely fixed because a freedom to rotate the spacecraft around the Z axis remains. This implies that the HGA azimuth is not fixed.
Figure~\ref{fig:HGApointing} shows the HGA azimuth and elevation evolution during transfer. It is clearly visible that the prohibited elevation range above $+25^\circ$ is violated for a large part of the transfer. This is a general feature of transfers to Earth-trailing orbits, because for the initial part of the transfer both Sun and Earth are in the same direction as seen from the spacecraft. This is can be seen in Figure~\ref{fig:transferTrajectory} (right). Various options for solving this problem are discussed, e.g. interruption of thrust arcs for Earth communications. For transfers to an Earth leading orbit the HGA elevation is usually not a problem.

\subsection{Sensitivity on the initial cartwheel clocking angle}
\label{subsec:clockingAngleSens}
As already mentioned in section~\ref{subsec:simultaneousOpti}, a key question for the design of the transfer trajectory is whether the operational orbit can be assumed as fixed for all launch dates or whether a custom operational orbit has to be designed as a function of the launch date. The MIDA, arm length and argument of perihelion are considered as (fixed) design parameters. The initial semi-major axis is mainly fixed by the maximum Earth distance constraint (cf. section~\ref{subsec:earthPerturb}) and the inclination parameter, $\delta$ is used to adjust the stability of the cartwheel geometry within narrow limits. Therefore, the only significant impact on the transfer~\dv~is expected from the initial cartwheel clocking angle, $\sigma_0$.

\begin{figure}
  \includegraphics[width=0.5\textwidth]{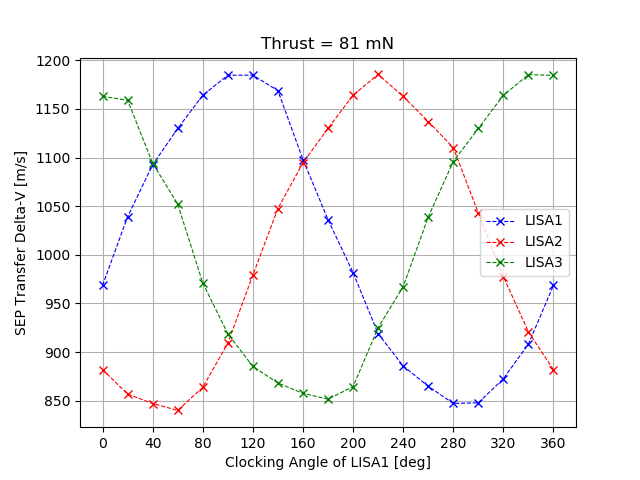}
	\includegraphics[width=0.5\textwidth]{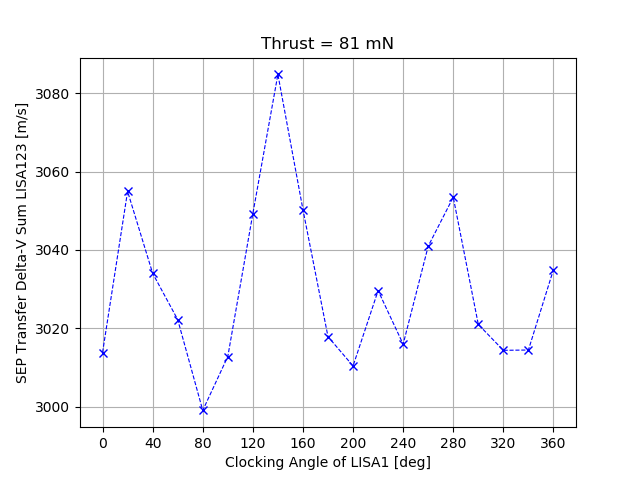}
\caption{Transfer~\dv~as function of the arrival orbit initial clocking angle. Individual LISA spacecraft (left) and sum of all LISA spacecraft (right).}    
\label{fig:clockingAngleSens}
\end{figure}

In order to assess the impact of this parameter on the transfer, a series of transfers similar to the one described in section~\ref{subsec:FebruaryLaunch}, have been computed, but varying the fixed value for the initial clocking angle. 
Figure~\ref{fig:clockingAngleSens} (left) shows the transfer \dv~of the three LISA spacecraft as a function of the initial clocking angle. A clear pattern is recognisable, which basically repeats after $120^\circ$ due to the symmetry of the formation. The difference between the largest and smallest maximum \dv~is about $80$ m/s. This shows that the clocking angle has to be optimized together with the transfer, in order to achieve the optimal \dv~for the propellant tank sizing. The plot also shows that an indication of the clocking angle being optimal is that the \dv~values of two LISA spacecraft is identical while the \dv~for the remaining spacecraft is below that value. Conversely, if all three spacecraft have different transfer \dv, it is an indication that the clocking angle has not been chosen optimally. For the transfer described in section~\ref{subsec:FebruaryLaunch} the clocking angle has been optimised along with the transfer and therefore the maximum transfer \dv~is close to $1100$ m/s.

Figure~\ref{fig:clockingAngleSens} (right) shows the sum of the \dv~of all three LISA spacecraft as a function of the clocking angle, which is relevant for computing the total wet mass on the launcher. Here the variation is less prominent, but still amounts to about 80 m/s for between the worst and the best clocking angle value. Note that the minima for the \dv~sum (right) don't occur at the same clocking angle as for the maximum \dv~value (left).

\subsection{Full year 2034 launch window}
\begin{figure}
  \includegraphics[width=0.99\textwidth]{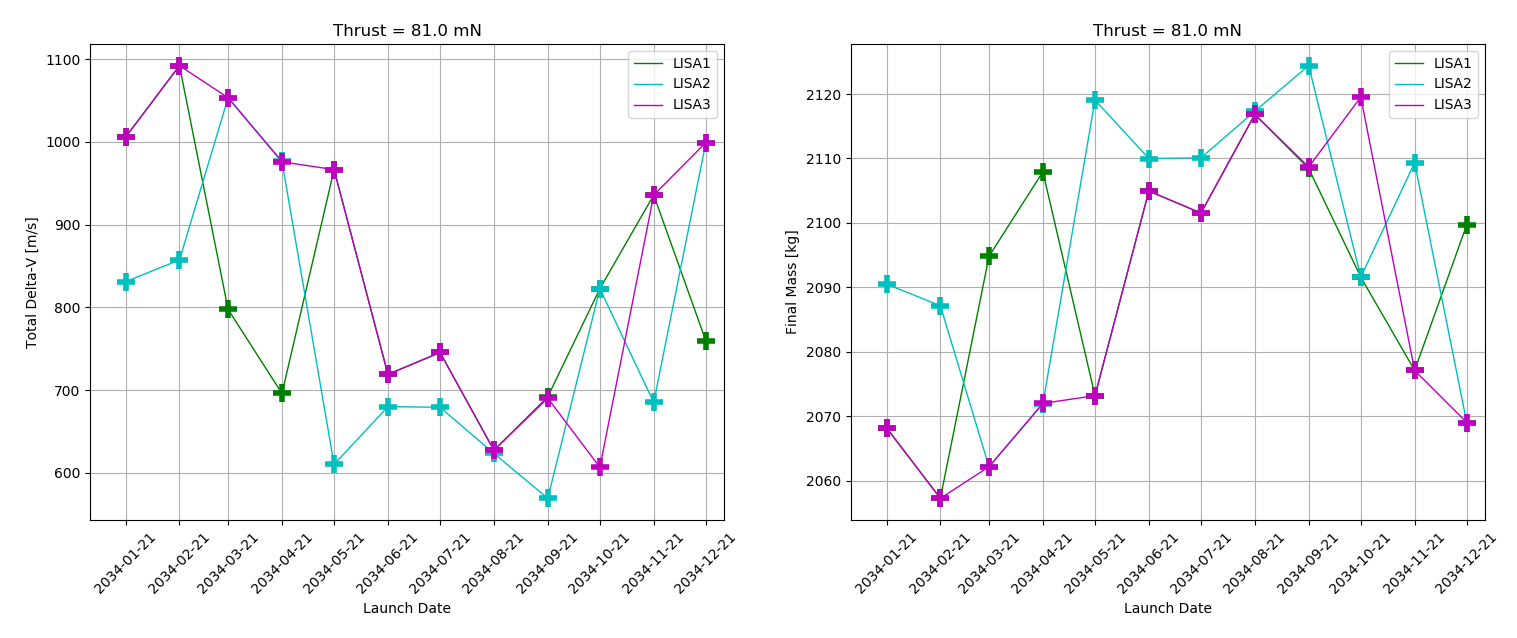}
\caption{SEP Transfer \dv~(left) and final mass (right) for 2034 MIDA$=-20^\circ$ clockwise. The crosses indicate actually computed cases. The lines are only drawn to guide the eye.}    
\label{fig:dv2034}
\end{figure}

Over the year a seasonal variation in the transfer \dv~is expected mainly for two reasons: the true anomaly of the Earth at departure and the relative geometry of the Earth equator to the ecliptic. 
The total \dv~is shown in Figure~\ref{fig:dv2034} (left). The sizing case happens in February which amounts to $1092$ m/s without margin (cf. section~\ref{subsec:FebruaryLaunch}). Note that the difference in \dv~between optimisation stages 1 and 3 (cf. section~\ref{subsec:simultaneousOpti}) was found to be below 12 m/s for all months in 2034. For the February launch the difference is 7 m/s. Therefore, even running only stage 1 yields an excellent estimate on the transfer \dv.
The transfer \dv~depends on how much the initial ecliptic RAAN has to be adjusted during the transfer of the individual spacecraft. Since the RAANs of the three spacecraft in the target orbit are separated by $120^\circ$, there can be significant differences between the three spacecraft. Note that the labelling LISA1/LISA2/LISA3 is arbitrary and can change between two months.

%The launch RAAN (in the EME2000 frame), shown in Figure~\ref{fig:raan2034} (left) more or less steadily increases over the year which implies a rather constant escape direction along the Earth velocity which is required to achieve a negative MIDA. Figure~\ref{fig:raan2034} (left) shows the ecliptic RAAN of the final operational orbit reached after the transfer. This parameter is a direct result of the operational orbit clocking angle optimization and thus every launch date %transfers to a different operational orbit.

%\begin{figure}
%  \includegraphics[width=0.99\textwidth]{SEP-MIDA-20deg-2034-RAAN.png}
%\caption{Launch RAAN (left) and science orbit RAAN (right) for 2034 MIDA$=-20^\circ$ clockwise.}    
%\label{fig:raan2034}
%\end{figure}

\begin{figure}
  \includegraphics[width=0.99\textwidth]{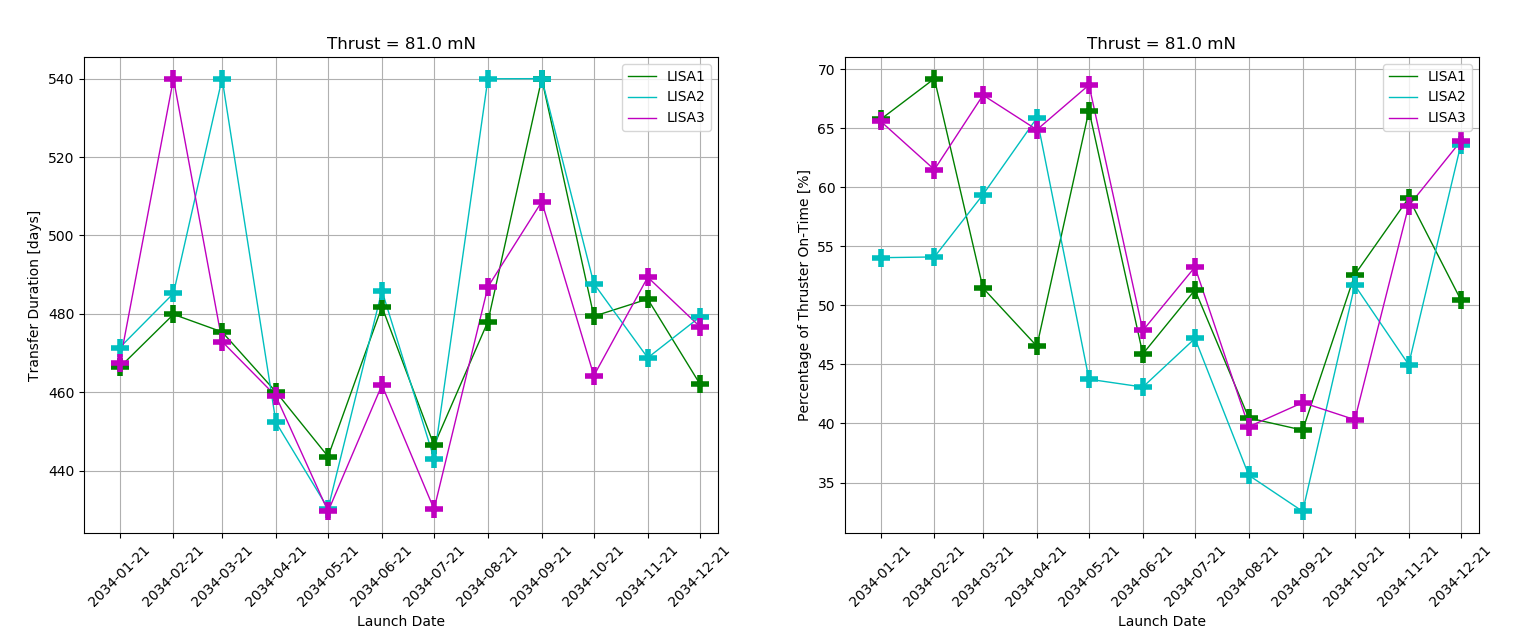}
\caption{SEP transfer duration (left) and percentage of thrust-on time (right) for 2034 MIDA$=-20^\circ$ clockwise.}    
\label{fig:dt2034}
\end{figure}

The total transfer duration varies between 440 days and 540 days over the year and between the three spacecraft as shown in Figure~\ref{fig:dt2034} (left). The total thrust-on time fraction, shown in Figure~\ref{fig:dt2034} (right), follows the same functional behaviour as the~\dv~and has a maximum of about $70\%$ in February. Reducing the thrust level below the $81$ mN is therefore expected to make these transfer opportunities more difficult to achieve.

\begin{figure}
  \includegraphics[width=0.99\textwidth]{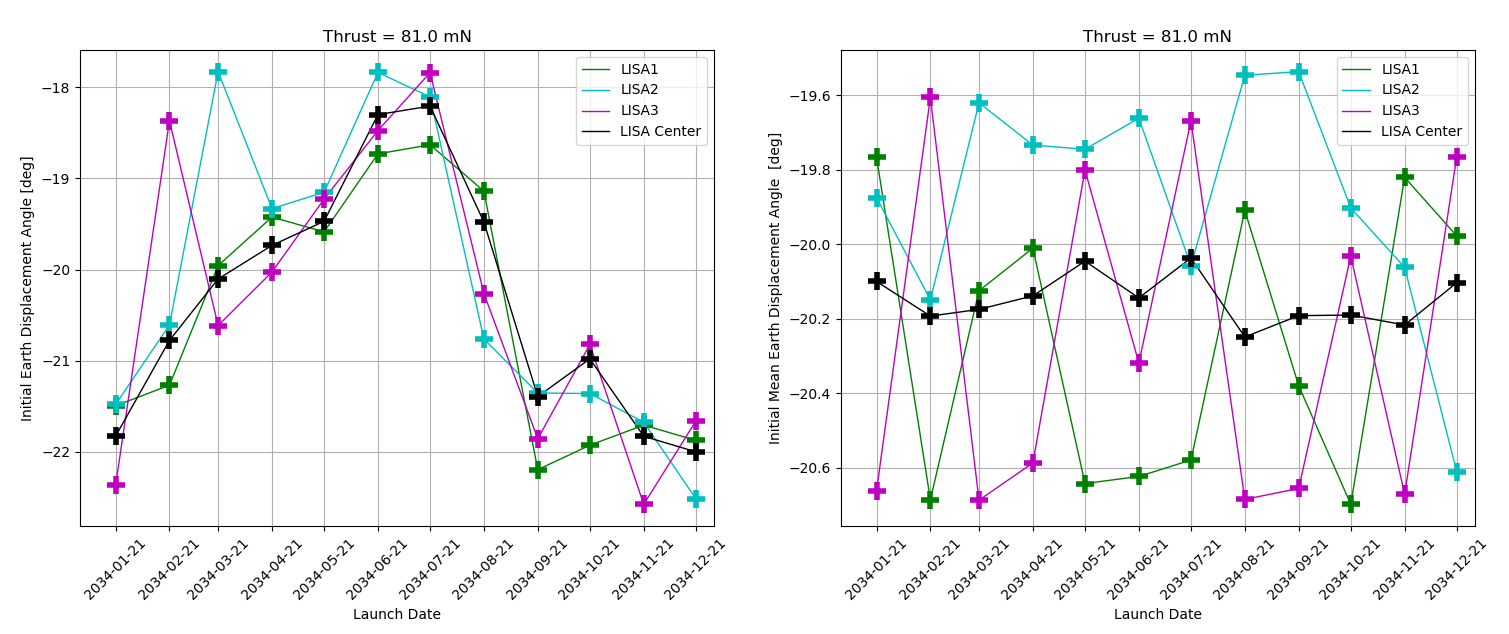}
\caption{Instantaneous initial Earth displacement angle (left) and MIDA (right) for 2034 MIDA$=-20^\circ$ clockwise.}    
\label{fig:mida2034}
\end{figure}

In order to illustrate the difference of the instantaneous initial Earth displacement angle versus the MIDA, these two parameters are plotted for the whole launch window in Figure~\ref{fig:mida2034}. As explained in section~\ref{subsec:MIDA}, the initial Earth displacement angle (left) experiences a variation of about $\pm 2^\circ$ over the year while the MIDA stays constant at $-20^\circ$. The slight deviation of the MIDA from $-20^\circ$ between the different launch opportunities stems from the fact that for the sake of simplicity the value of MIDA$=-20^\circ$ was imposed 400 days after launch for all launch dates. Since the transfer duration varies between the different launch opportunities, the actually achieved MIDA has a (negligible) variation.

\begin{figure}
  \includegraphics[width=0.99\textwidth]{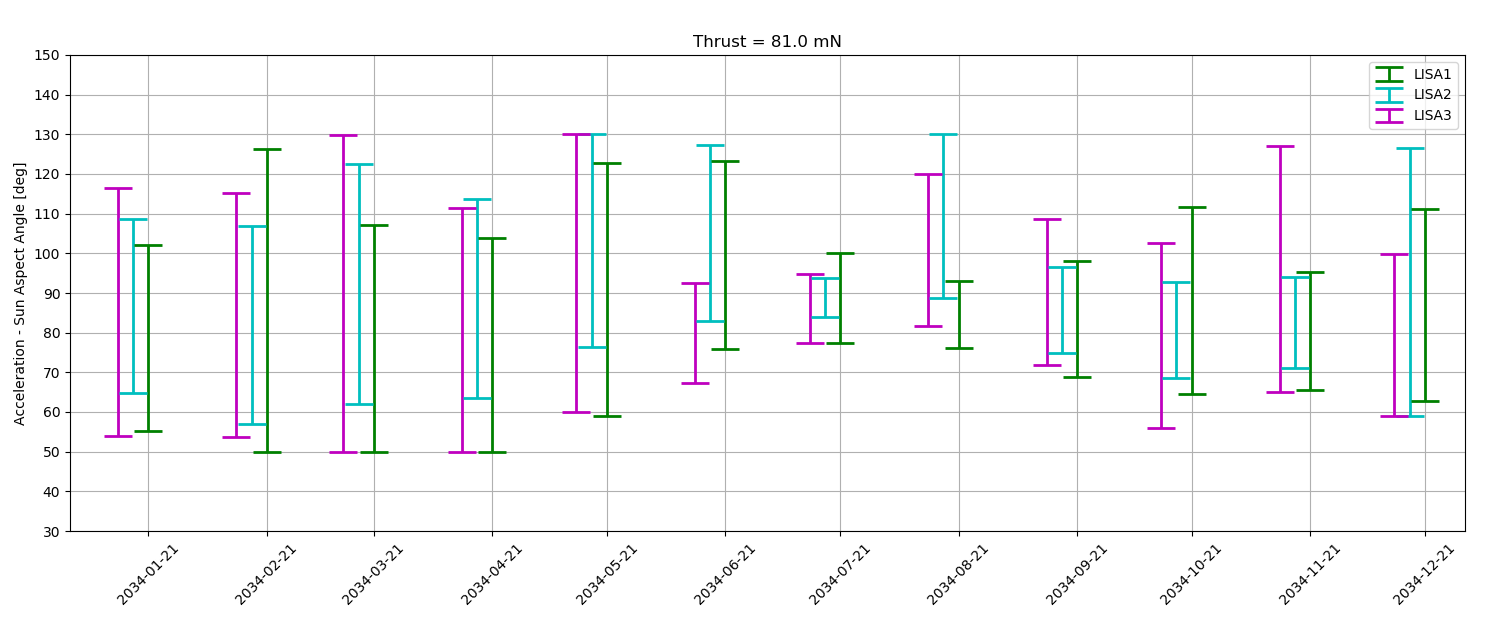}
\caption{Ranges of Sun aspect angles of the acceleration vector for SEP transfers in 2034 MIDA$=-20^\circ$ clockwise.}    
\label{fig:saa2034}
\end{figure}

The ranges of Sun aspect angles are always within the allowed range between $50^\circ$ and $130^\circ$, as shown in Figure~\ref{fig:saa2034}. In most cases either the lower or the upper constraint is active.

\begin{figure}
  \includegraphics[width=0.49\textwidth]{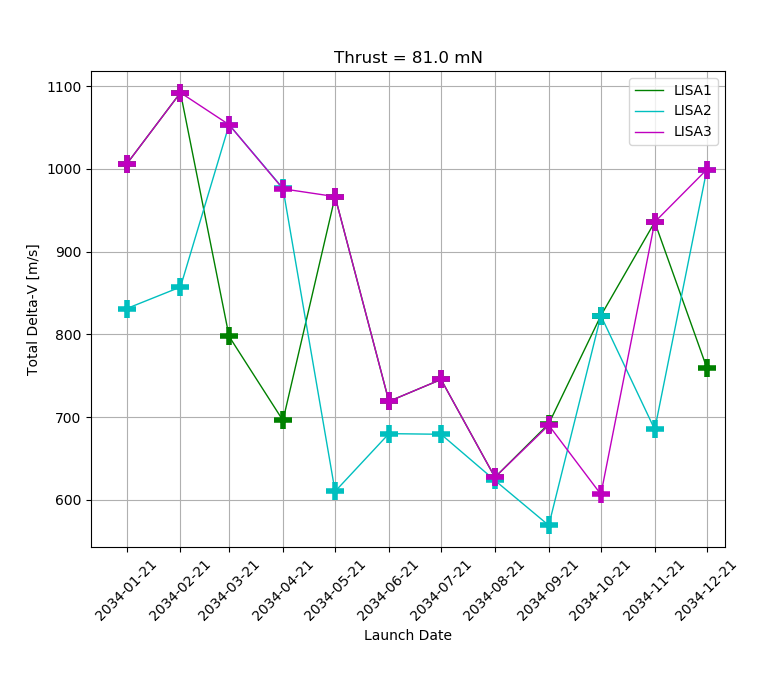}
	\includegraphics[width=0.49\textwidth]{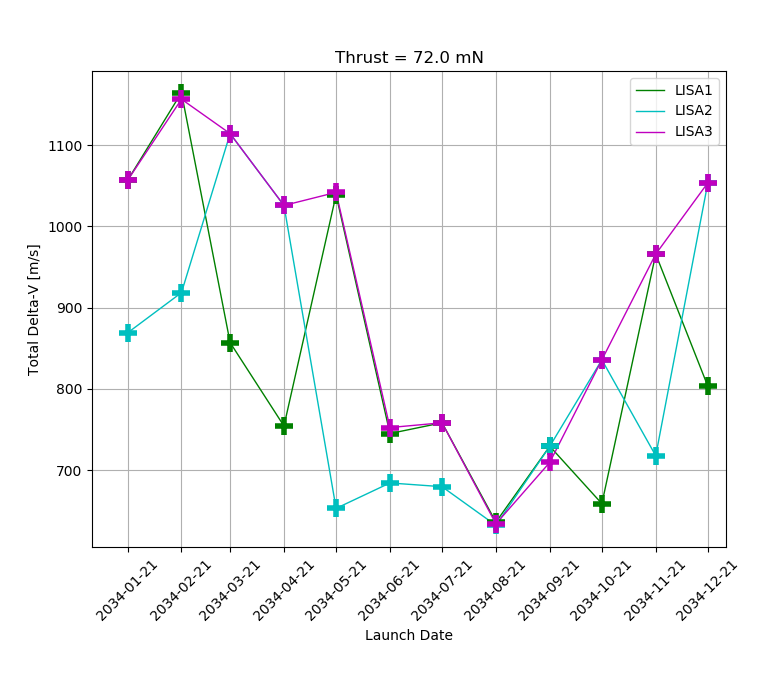}\\
	\includegraphics[width=0.49\textwidth]{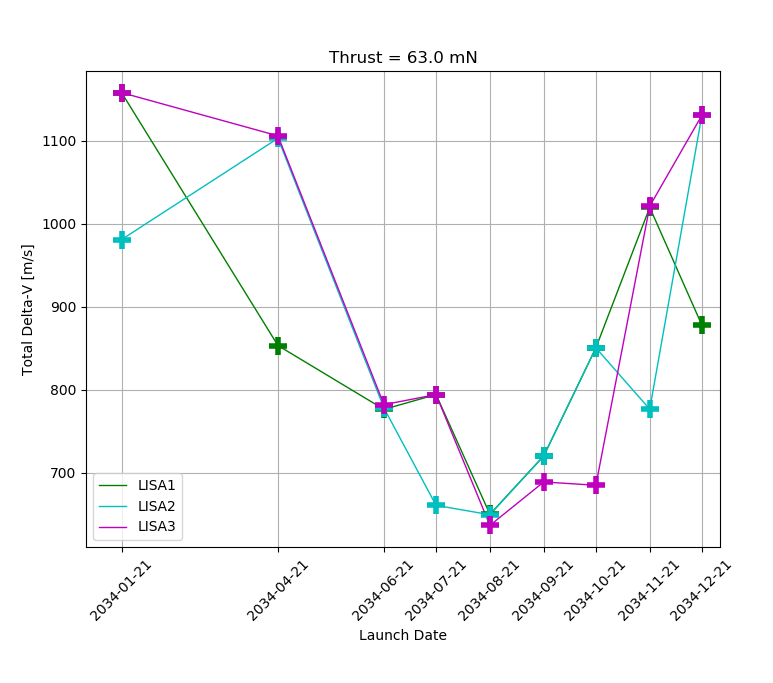}
	\includegraphics[width=0.49\textwidth]{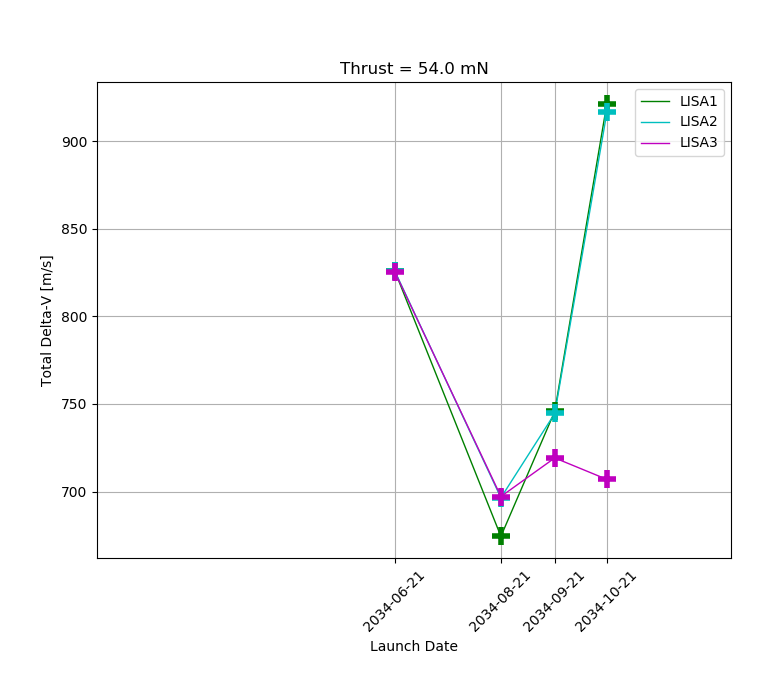}
\caption{Total \dv~for different thrust level values for 2034 MIDA$=-20^\circ$ clockwise.}    
\label{fig:dv81mN-54mN}
\end{figure}

In order to assess the dependency of the \dv~on the thrust capability of the spacecraft, several cases with a reduced thrust level have been analysed. The results for (effective) thrust levels of 81 mN, 72 mN, 64 mN and 54 mN are shown next to each other in Figure~\ref{fig:dv81mN-54mN}. Besides of increasing the required \dv, the effect of the lower thrust is also to reduce the number of launch opportunities. For 54 mN it is quite difficult to achieve convergence at all. Therefore, this option is not very robust and thus not recommended.

So far only transfers to an Earth-trailing orbit have been discussed. The analysis for a Earth-leading orbit has also been conducted but will not be presented in detail here for the sake of brevity. The main difference in the transfer \dv~is the change in the seasonal variation: The sizing case for a MIDA of $+20^\circ$ occurs in August instead of February and amounts to 1253.3 m/s. This fact can be exploited in order to reduce the overall sizing \dv~by combining launches to trailing and leading orbits depending on the launch month: In the Summer months a launch to a trailing orbit is preferable while in the Winter months a leading orbit yields the lower transfer \dv~\cite{JoffreLeadingTrailing}.

\section{Navigation and insertion accuracy}
\label{sec:navigation}
This section is presenting all the analyses related to orbit navigation and guidance of the LISA spacecraft during transfer and cartwheel orbit insertion. A key question to be answered here is: how precisely can the three LISA spacecraft be inserted into the cartwheel formation and how does the error impact the stability of the cartwheel? The first part of that question is going to be answered by the presented navigation analysis, the second part by a Monte-Carlo analysis that uses the output dispersion matrix from the navigation as an input.
The idea of a navigation analysis is to simulate all ground-based (and potentially inter-spacecraft) measurements as well as orbit correction manoeuvres that will take place during the operations phase. Generally speaking, there are two key outputs of such an analysis that quantify the orbit accuracy that can be reached:
\begin{enumerate}
\item Knowledge: the difference between the true spacecraft state and the estimated spacecraft state. This is mainly a function of the measurement accuracy and geometry of the ground-based Range and Doppler observations, but is also influenced by noise processes like the SEP acceleration or SRP.
\item Dispersion: the difference between the true spacecraft state and the reference spacecraft state. This is mainly a result of manoeuvre execution errors, but is also influenced by the state knowledge: the dispersion can never get lower than the knowledge at a given epoch, because the precision to which manoeuvres are computed by Flight Dynamics can never exceed the state knowledge on which these computations are based.
\end{enumerate}

Both figures are generally described by covariance matrices and are typically shown in the local orbital frame of the spacecraft.

A navigation analysis can be broken down into two parts: the covariance analysis, which mainly deals with the knowledge and the guidance analysis which mainly deals with the dispersion. Since, however, both are inter-linked, iterations between the two are needed until convergence of all covariance matrices and the manoeuvre budget is reached.

All analyses in this chapter are based on the reference trajectory with launch in March 2034. Similar results are expected for any other transfer. 

\subsection{Covariance analysis assumptions}
\label{subsec:covAssumptions}

The covariance analysis uses a Square-Root-Information Filter (SRIF) to process all available observation types and compute the covariance matrix of all estimated parameters. Some uncertain parameters in the problem are not estimated (i.e. their a-priori covariance is chosen not to be affected by the observations processing), but their uncertainty still needs to be taken into account. These parameters are called consider parameters. A summary of all uncertain parameters is given in Table~\ref{tab:covAssumptions}.

\begin{table}
\caption{Summary of all uncertain parameters that appear in the covariance analysis.}
\label{tab:covAssumptions}     
\begin{tabular}{lll}
\hline\noalign{\smallskip}
Parameter                                     & A-priori $1-\sigma$                     & Status    \\
\noalign{\smallskip}\hline\noalign{\smallskip}
LISA 1,2,3 states                             &	$1000$ km, $3$ m/s spherical	          & estimated  \\
Guidance manoeuvres                           &	Output from guidance analysis	          & estimated  \\
SEP acceleration (white) noise                &	$1 \%$ magnitude, $0.5^\circ$ direction	& estimated  \\
SRP (white) noise                             &	$1 \%$ of computed magnitude	          & estimated  \\
Generic non-gravitational acceleration noise  &	$5 \cdot 10^{-12}$ km/s$^2$             & estimated  \\
Ground station position bias	                & $30$ cm spherical	                      & considered \\
Range observation bias	                      & $10$ m	                                & considered \\
\noalign{\smallskip}\hline
\end{tabular}
\end{table}

Observations are used by the filter to improve the knowledge of the estimated parameters. Currently, only ground-based Range and Doppler measurements are assumed. All observation assumptions are summarized in Table~\ref{tab:navObservations}.

\begin{table}
\caption{Summary of all observation assumptions in the covariance analysis.}
\label{tab:navObservations}     
\begin{tabular}{lllll}
\hline\noalign{\smallskip}
Observation  & Random noise     & Measurement interval & Data cut-off    &  Ground Station  \\
\noalign{\smallskip}\hline\noalign{\smallskip}
Range   &	2 m	                            &  1 hour	& 4 days  &	Cebreros DSA \\
Doppler & 0.3 mm/s              &  10 min	& 4 days  &  Cebreros DSA \\
        &  (60 sec count time)	&       	&         &   \\
\noalign{\smallskip}\hline
\end{tabular}
\end{table}

All other assumptions related to the covariance analysis are summarized in Table~\ref{tab:navOther}.

\begin{table}
\caption{Other assumptions related to covariance analysis.}
\label{tab:navOther}     
\begin{tabular}{ll}
\hline\noalign{\smallskip}
Quantity    &    Value  \\
\noalign{\smallskip}\hline\noalign{\smallskip}
Minimum spacecraft elevation above the ground station horizon	&   15 deg  \\
Orbit determination arc length	                              &   14 days, sliding window \\
Ground station contact time per 7 days per spacecraft	        &    8 hours. \\
                                                              &  Scheduling based on priority \\
																															&   in the order \\
																															&   LISA1, LISA2, LISA3 \\
\noalign{\smallskip}\hline
\end{tabular}
\end{table}

\begin{figure}
  \includegraphics[width=0.99\textwidth]{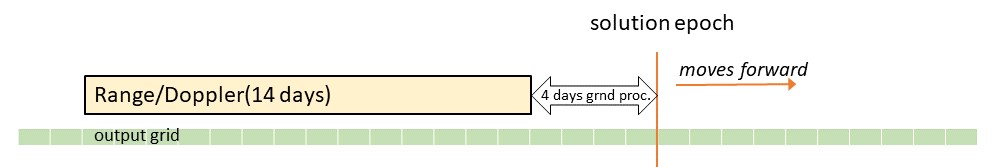}
\caption{Illustration of orbit determination process using a sliding window.}    
\label{fig:slidingWindow}
\end{figure}

In order to simulate the orbit determination process done in operations as closely as possible, a sliding window is used as an orbit determination arc. This means for every solution epoch measurements 14 days before the corresponding data cut-off epoch are processed (4 days for Range and Doppler). The obtained covariance matrix is then mapped forward to the solution epoch. The process is repeated for the considered range of solution epochs and the resulting knowledge evolution is plotted in a graph. This procedure is illustrated in Figure~\ref{fig:slidingWindow}.

\subsection{Guidance analysis assumptions}
\label{subsec:guidanceAssumptions}
The purpose of the guidance analysis is to determine to which extent the spacecraft state dispersion can be reduced by manoeuvres using the state estimation from the covariance analysis as input. To this end, guidance manoeuvres are scheduled that target the reference state at a pre-defined epochs. 
For simplicity, the current implementation of the analysis assumes impulsive guidance manoeuvres although LISA is only equipped with SEP. This approach is justified since the resulting \dv~per manoeuvre is expected to be small, which is confirmed a-posteriori. Therefore, these guidance manoeuvres can also be performed by the SEP with roughly the same \dv~by simply extending the burn duration. During the trajectory design a margin of $10 \%$ on the available thrust magnitude has been taken in order to accommodate (amongst other things) such guidance manoeuvres even if they happen during a deterministic thrust arc.

Table~\ref{tab:guidanceAssumptions} summarises the general assumptions for the guidance analysis.
\begin{table}
\caption{Assumptions for guidance analysis.}
\label{tab:guidanceAssumptions}     
\begin{tabular}{ll}
\hline\noalign{\smallskip}
Quantity    &    Value  \\
\noalign{\smallskip}\hline\noalign{\smallskip}
Initial dispersion for cartwheel insertion	       &    200 km, 2 m/s spherical \\
Initial epoch	                                     &    2035-07-26 00:00:00   \\
Guidance manoeuvre mechanisation error $1-\sigma$	 &    $1 \%$ magnitude, $0.5^\circ$ direction, $1$ mm/s minimum \\
\noalign{\smallskip}\hline
\end{tabular}
\end{table}

For the cartwheel insertion analysis the Trajectory Correction Manoeuvre (TCM) schedule is centred on the cartwheel arrival date, which is defined at the end of the last thrust arc. Since the last thrust arc ends at a different epoch for all three spacecraft, we use the latest date of the three as a reference, which is 2035-09-12 12:00:00 for the reference trajectory used here. In order to accommodate enough time for ground station observations and ground processing between the manoeuvres, they are scheduled every 14 days as shown in Table~\ref{tab:manoeuvreSchedule}. For simplicity, currently all three spacecraft are assumed to execute the TCMs at the same epochs.

\begin{table}
\caption{TCM schedule for cartwheel insertion.}
\label{tab:manoeuvreSchedule}     
\begin{tabular}{lll}
\hline\noalign{\smallskip}
Manoeuvre name    &    Manoeuvre epoch    & Target and epoch  \\
\noalign{\smallskip}\hline\noalign{\smallskip}
TCM1  &	2035-08-15 12:00:00	& Position at 2035-09-12 12:00:00  \\
TCM2  &	2035-08-29 12:00:00	& Position at 2035-09-12 12:00:00  \\
TCM3  &	2035-09-12 12:00:00	& Position at 2035-10-10 12:00:00  \\
TCM4  &	2035-09-26 12:00:00	& Position at 2035-10-10 12:00:00  \\
TCM5  &	2035-10-10 12:00:00	& Velocity at 2035-10-10 12:00:00  \\
\noalign{\smallskip}\hline
\end{tabular}
\end{table}

\begin{figure}
  \includegraphics[width=0.99\textwidth]{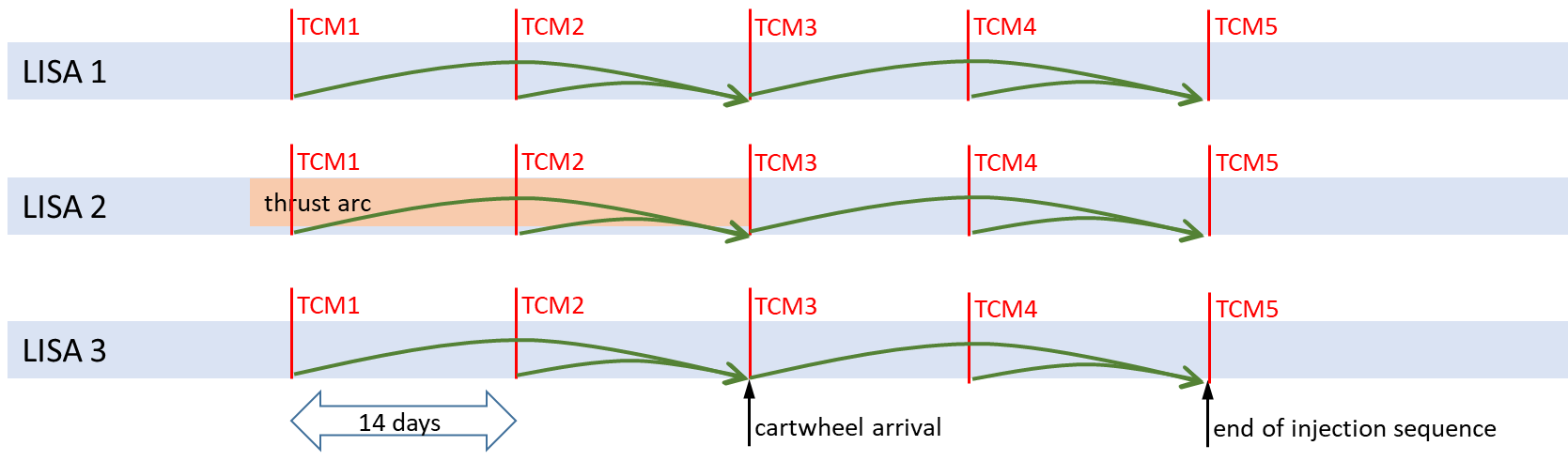}
\caption{Illustration of cartwheel insertion TCM strategy. The green arrows indicate the targeting point of the TCMs. TCM5 targets the velocity vector at its epoch.}    
\label{fig:guidanceSchedule}
\end{figure}

Figure~\ref{fig:guidanceSchedule} shows a graphical representation of the manoeuvre schedule for the three spacecraft, also showing the SEP thrust arcs. This illustrates that only for LISA2 the first two TCMs happen during an SEP thrust arc. Due to SEP noise these manoeuvres are expected to be much less efficient than for LISA1 and LISA3 where they happen during a coast arc. In that sense LISA2 will represent the sizing/worst case in terms of insertion accuracy. On the other hand, the analysis for LISA1 and LISA3 will show how much can be gained by not doing any guidance manoeuvres during thrust arcs for the cartwheel insertion.
The targeting strategy is based on the idea that in order to match both position and velocity of the cartwheel state at the end of the sequence, at least two TCMs are needed: the first one will match the position only and the second one will match the velocity. However, due to manoeuvre execution error, the match will never be exact. Therefore, two manoeuvres are assumed for each target point, a larger and a smaller one. Finally, TCM5 is an additional manoeuvre to clean up any residual velocity dispersion.

\subsection{Navigation results}
\label{subsec:navResults}
\begin{figure}
  \includegraphics[width=0.99\textwidth]{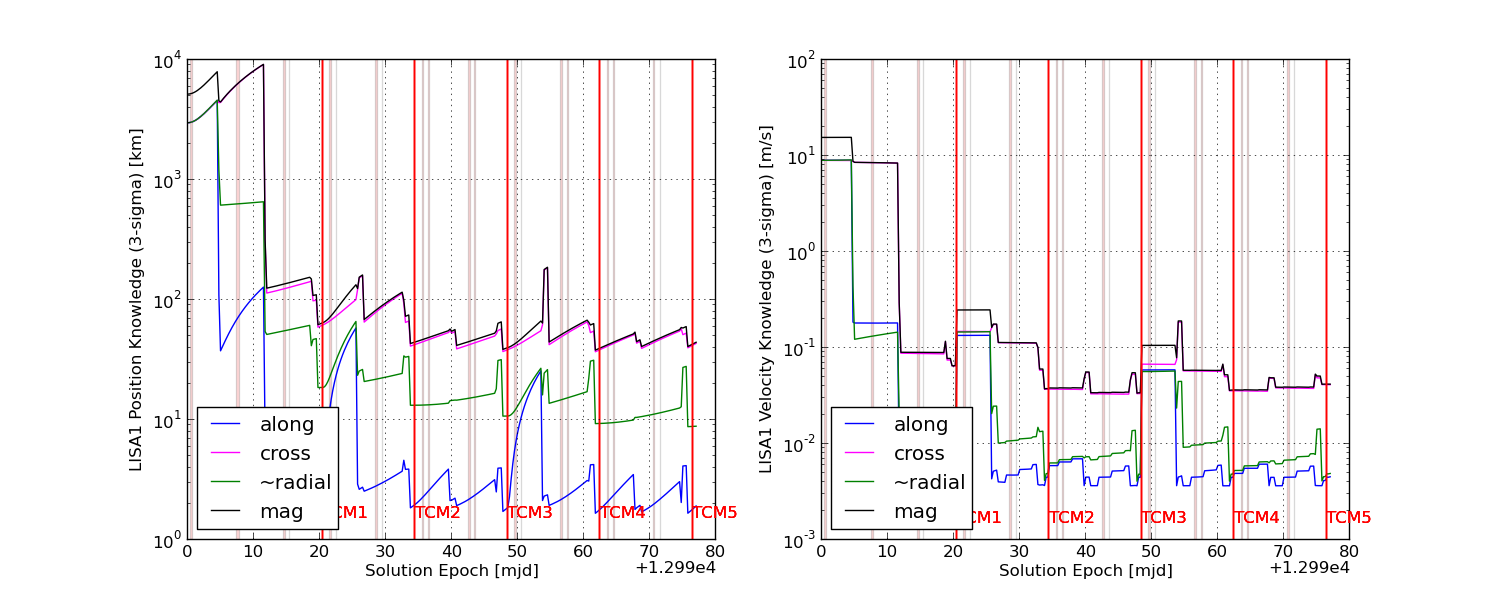}
	\includegraphics[width=0.99\textwidth]{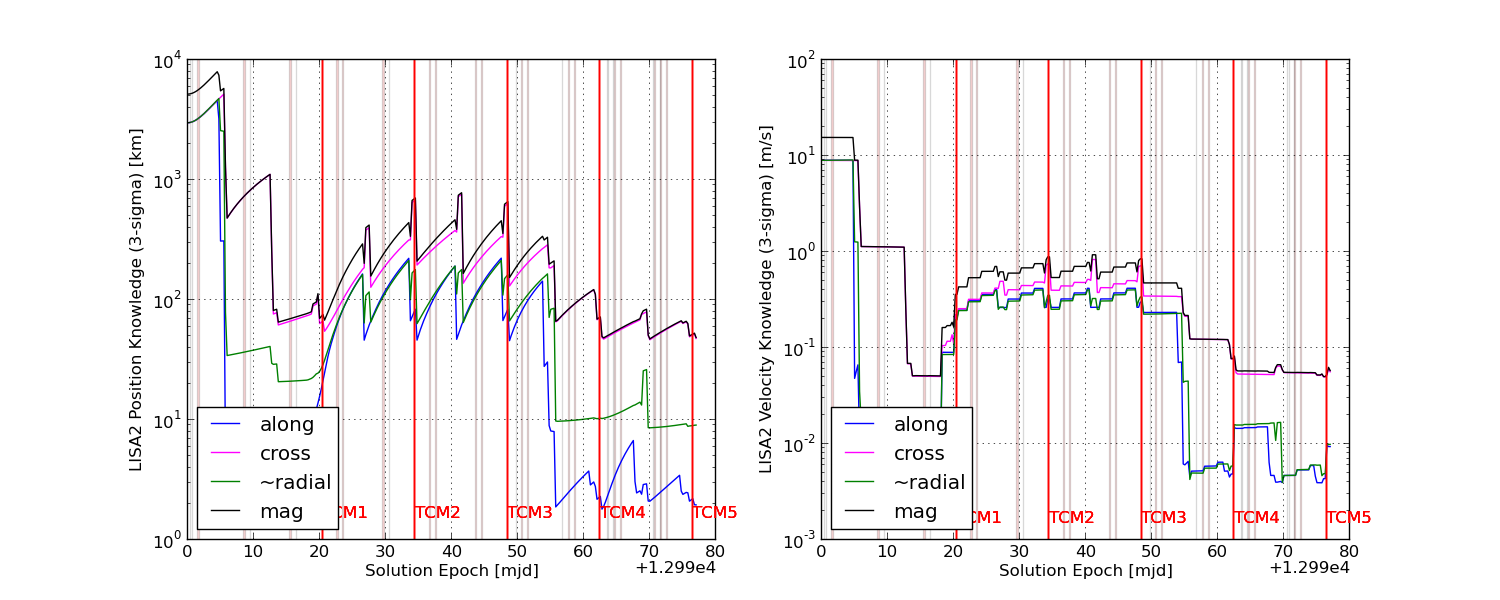}
	\includegraphics[width=0.99\textwidth]{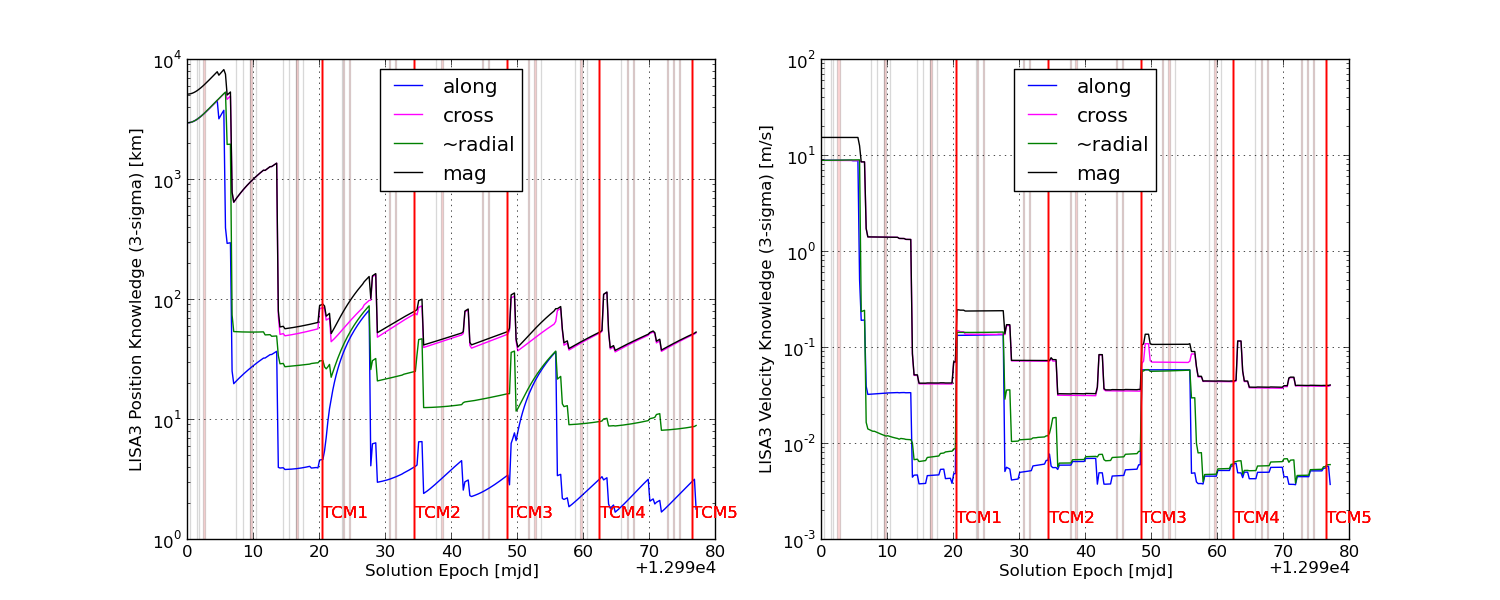}
\caption{Knoweldge covariance evolution of all three LISA spacecraft for Range and Doppler (8 hours per week per s/c).}    
\label{fig:knowledge}
\end{figure}

Figure~\ref{fig:knowledge} shows the knowledge evolution of both position and velocity in the local orbital frame for all three LISA spacecraft ($3-\sigma$ values). The initially large covariance is reduced as soon measurements are taken (indicated by shaded vertical bars). This happens with a latency of 4 days due to the data cut-off. When no measurements are taken, the knowledge slowly degrades owing to the natural dynamics as well as the system noise. The impact of the noise is particularly strong for LISA2 which is the only spacecraft that has an SEP thrust arc in the considered time period. Jumps in the velocity knowledge are also caused by the guidance manoeuvres, TCM1 - TCM5. The along-track component in most cases is the one with the best knowledge. This is because of the measurement geometry: in a trailing/leading orbit ground-based Range and Doppler always measure the along-track component directly. The signature in an extended Doppler measurement arc imparted by the Earth's rotation also measures the plane-of-sky components. However, the vertical component (= cross-track) is not measured well if the spacecraft is located close to zero declination w.r.t. the Earth's equator, which is the case here. This is generally known as the zero declination problem. A way to significantly improve the plane-of-sky resolution, is to use $\Delta$DOR measurements. As will be shown in section~\ref{subsec:stability}, the along-track knowledge is the driving one for the cartwheel stability. Therefore, $\Delta$DOR measurements are not required here.

\begin{figure}
  \includegraphics[width=0.99\textwidth]{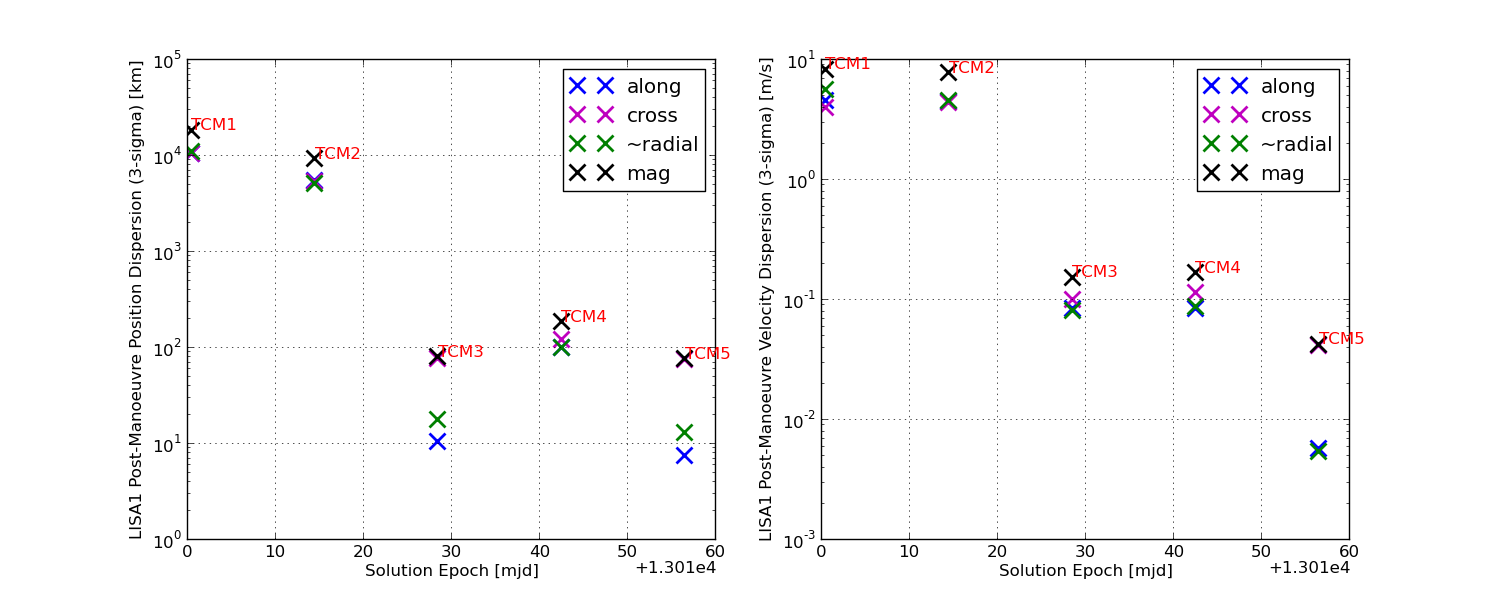}
	\includegraphics[width=0.99\textwidth]{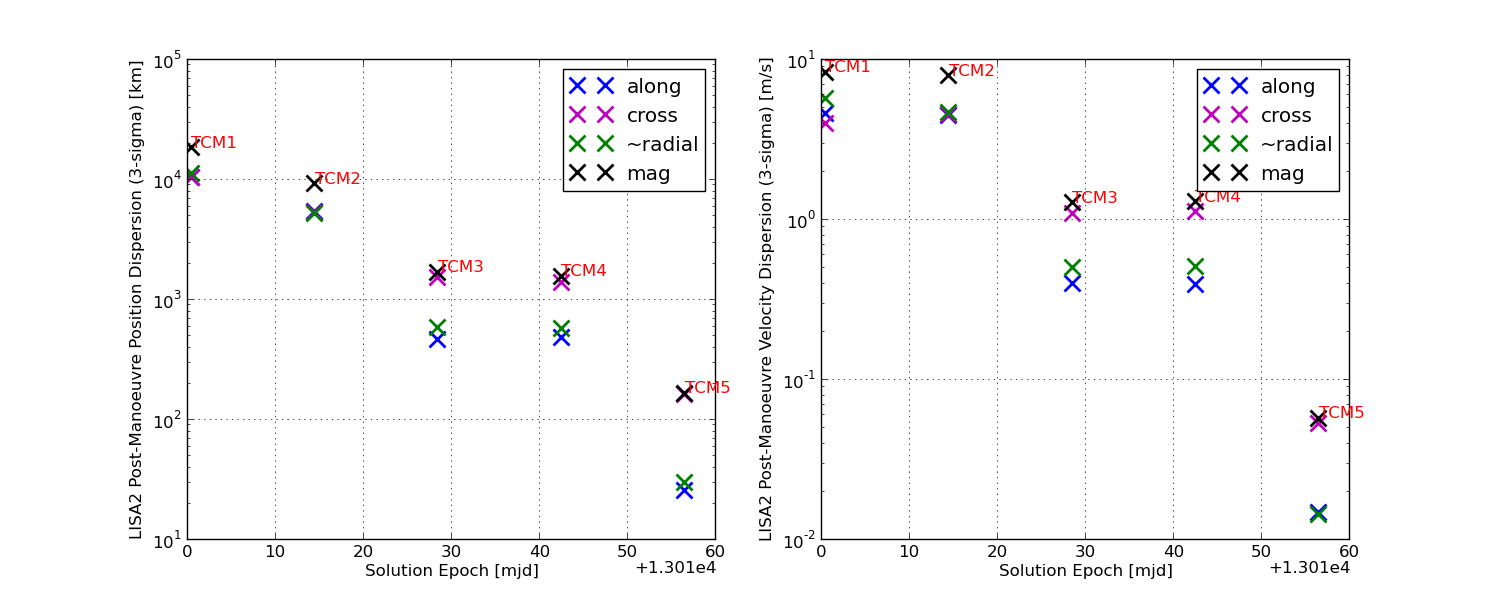}
	\includegraphics[width=0.99\textwidth]{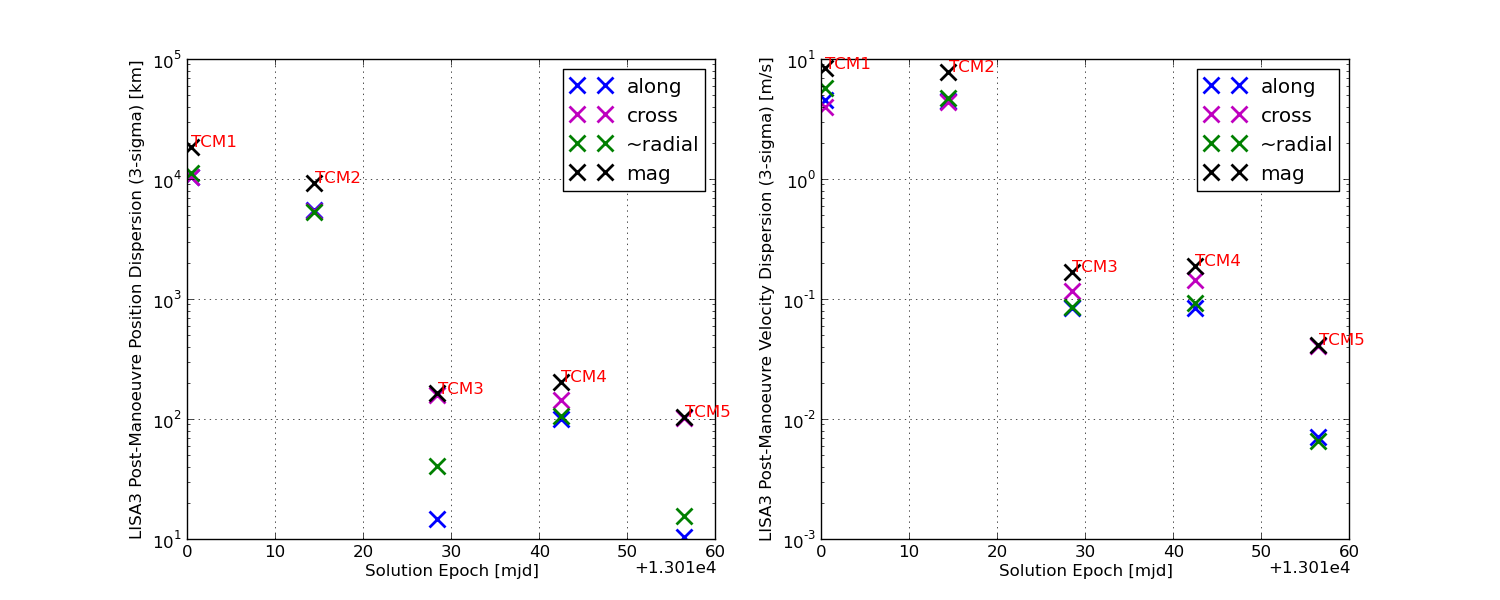}
\caption{Post-TCM dispersion of all three LISA spacecraft for Range and Doppler (8 hours per week per s/c).}    
\label{fig:dispersion}
\end{figure}

The knowledge covariance is used to determine guidance manoeuvres in order to reduce the initially large dispersion. The post-manoeuvre dispersion resulting from the guidance analysis are shown in Figure~\ref{fig:dispersion}. One can see how the dispersion is gradually improved as more guidance manoeuvres are executed. The state components with the best knowledge also end up with the least dispersion.

Finally, Table~\ref{tab:stochasticDV1},~\ref{tab:stochasticDV2} and~\ref{tab:stochasticDV3} present the stochastic manoeuvre budget resulting from the guidance analysis. It is clear that TCM1 and TCM3 are the largest manoeuvres because they are the first manoeuvres for each of the two target points. TCM2, TCM4 and TCM5 are follow-up manoeuvres which merely clean up the mechanisation error of the previous manoeuvres. 
For the fuel tank sizing the $99 \%$ CL value shall be taken which amounts to 18.3 m/s per spacecraft. This value is still preliminary since the initial dispersion used here was guessed and does not originally come from the launcher dispersion and an end-to-end navigation analysis. This end-to-end navigation analysis shall also use continuous SEP guidance manoeuvres. Possibly, part of the corrections can then be absorbed into the deterministic thrust arcs.

\begin{table}
\caption{Stochastic manoeuvre budget for LISA1}
\label{tab:stochasticDV1}     
\begin{tabular}{lllllll}
\hline\noalign{\smallskip}
Maneuver  	&   Mean (m/s)	&   Std (m/s)	&  $90\%$ (m/s)   &  	$95 \%$ (m/s)	  & $99\%$ (m/s)   \\
\noalign{\smallskip}\hline\noalign{\smallskip}
TCM1 	&  5.555  &	 2.348  & 8.653	  &  9.723 &	12.046 \\
TCM2	&  0.211	&  0.103	&  0.345	&  0.399 &	0.527\\
TCM3	&  2.351	&  0.983	&  3.675	&  4.094 &	4.965\\
TCM4	&  0.092	&  0.044	&  0.149	&  0.171 &	0.224\\
TCM5	&  0.051	&  0.024	&  0.084	&  0.096 &	0.124\\
\noalign{\smallskip}\hline\noalign{\smallskip}
TOTAL	&  8.259	&  3.372	&  12.773 &	14.264 &	17.46\\
\noalign{\smallskip}\hline
\end{tabular}
\end{table}

\begin{table}
\caption{Stochastic maneuver budget for LISA2}
\label{tab:stochasticDV2}     
\begin{tabular}{lllllll}
\hline\noalign{\smallskip}
Maneuver  	&   Mean (m/s)	&   Std (m/s)	&  $90\%$ (m/s)   &  	$95 \%$ (m/s)	  & $99\%$ (m/s)   \\
\noalign{\smallskip}\hline\noalign{\smallskip}
TCM1 &	5.56 &	2.372 &	8.662 &	9.777 &	11.985\\
TCM2 &	0.493 &	0.25 &	0.837 &	0.967 &	1.24\\
TCM3 &	2.451 &	1.032 &	3.836 &	4.301 &	5.173\\
TCM4 &	0.621 &	0.338 &	1.095 &	1.276 &	1.65\\
TCM5 &	0.378 &	0.213 &	0.673 &	0.789 &	1.029\\
\noalign{\smallskip}\hline\noalign{\smallskip}
TOTAL	&  9.502 & 3.383 &	13.998 &	15.438 & 18.324\\
\noalign{\smallskip}\hline
\end{tabular}
\end{table}

\begin{table}
\caption{Stochastic maneuver budget for LISA3}
\label{tab:stochasticDV3}     
\begin{tabular}{lllllll}
\hline\noalign{\smallskip}
Maneuver  	&   Mean (m/s)	&   Std (m/s)	&  $90\%$ (m/s)   &  	$95 \%$ (m/s)	  & $99\%$ (m/s)   \\
\noalign{\smallskip}\hline\noalign{\smallskip}
TCM1 &	5.612 &	2.412 &	8.865 &	9.984 &	12.159\\
TCM2 &	0.22 &	0.106 &	0.357 &	0.411 &	0.547\\
TCM3 &	2.369 &	0.998 &	3.707 &	4.165 &	4.998\\
TCM4 &	0.096 &	0.046 &	0.157 &	0.182 &	0.232\\
TCM5 &	0.057 &	0.028 &	0.096 &	0.11 &	0.14\\
\noalign{\smallskip}\hline\noalign{\smallskip}
TOTAL	& 8.355	& 3.45	& 13.0	& 14.539& 	17.592\\
\noalign{\smallskip}\hline
\end{tabular}
\end{table}

\subsection{Cartwheel stability under insertion dispersions}
\label{subsec:stability}
A key aspect for the LISA operational orbit design is the question whether the formation stability can be maintained given the insertion accuracies. The current section uses the dispersion matrices at the end of the insertion sequence that are a product of the navigation analysis, section~\ref{subsec:navResults}, in order to do a Monte-Carlo analysis. The three spacecraft states after the last TCM are sampled and propagated for 10 years. 10,000 samples are used.

\begin{figure}
  \includegraphics[width=0.99\textwidth]{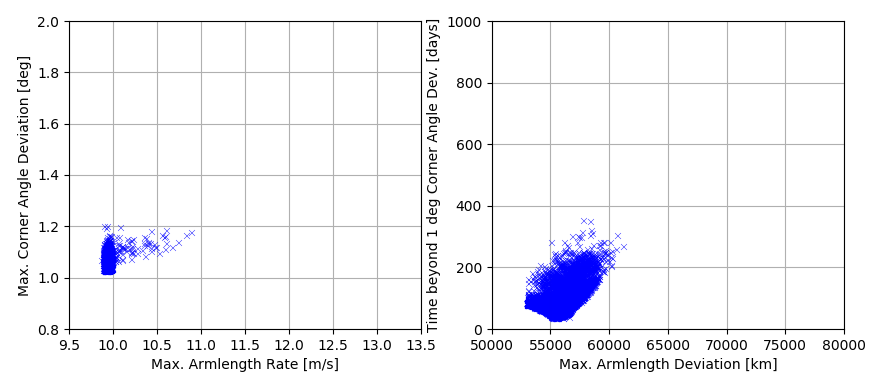}
\caption{Formation stability for MIDA$=-20^\circ$ Range, Doppler case.}    
\label{fig:cartwheel-dispersion-noSG}
\end{figure}

The resulting dispersions are presented in Figure~\ref{fig:cartwheel-dispersion-noSG}. The left plot shows the distribution of maximum arm length rate vs. maximum corner angle deviation from $60^\circ$. The distribution of the maximum arm length deviation from $2.5\cdot 10^6$ km vs. the time spent in a formation with at least one corner angle exceeding the $60^\circ\pm 1^\circ$ range is shown on the right. Apparently, the deviation of the samples is very well controlled for the key parameters (arm length rate and corner angle deviation). The duration spent in a formation fulfilling the corner angle requirement is below one year in most cases. The maximum Earth distance is not affected at all by the insertion inaccuracies, as expected.

A more quantitative summary of the analysis is given in Table~\ref{tab:stabilityStatistics}. The $99 \%$ C.I. quantities are recommended to be used as reliable benchmarks of the expected variation. These don't differ very much from the requirements of $10$ m/s arm length rate and $1^\circ$ corner angle deviation. Note that the computation of time spent in a formation where at least one of the three corner angles is outside the $60^\circ\pm 1^\circ$ range is based on a time discretisation step of 2 days. Therefore, the given number has an intrinsic error coming from the discretisation. The order of this error is 2 days times the number of intervals where a corner angle exceeds the allowed range.

\begin{table}
\caption{Formation stability statistics for MIDA$=-20^\circ$ Range, Doppler case.}
\label{tab:stabilityStatistics}     
\begin{tabular}{lll}
\hline\noalign{\smallskip}
          &   Minimum    & Maximum \\
\noalign{\smallskip}\hline\noalign{\smallskip}
Overall Min./Max. Armlength [km]         & 2438701.1 & 2536232.5\\
$99 \%$ C.I. of Min./Max. Armlength [km] & 2441140.4 & 2531975.9\\
Overall Min./Max. Armlength Rate [m/s] &    -10.90 &     10.01\\
$99 \%$ C.I. of Min./Max. Armlength Rate [m/s] &     -9.94 &      9.99\\
Overall Min./Max. Corner Angle [deg] &    58.806 &    61.199\\
$99 \%$ C.I. of Min./Max. Corner Angle [deg] &    58.890 &    61.103\\
$99 \%$ C.I. Time spent beyond $60^\circ\pm 1.0^\circ$ [days] &       230 &  \\
\noalign{\smallskip}\hline
\end{tabular}
\end{table}

\subsection{Cartwheel stability including unknown self-gravity accelerations}
\label{subsec:stabilitySelfGravity}
\begin{figure}
  \includegraphics[width=0.99\textwidth]{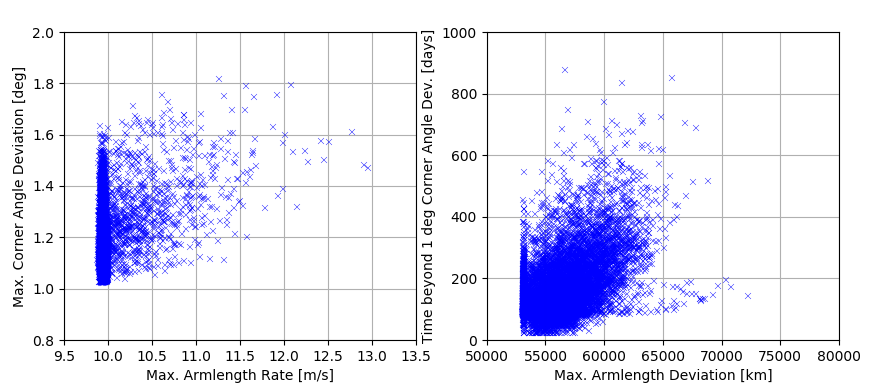}
\caption{Formation stability for MIDA$=-20^\circ$ Range, Doppler case with self-gravity accelerations.}    
\label{fig:cartwheel-dispersion-withSG}
\end{figure}

The picture changes if one adds the uncertainty from the acceleration component due to the spacecraft self-gravity. The effect of a known deterministic self-gravity acceleration has been analysed in section~\ref{subsec:selfGravity}. In reality, the magnitude and direction of the self-gravity acceleration will depend on the spacecraft configuration, component placement tolerances and fuel depletion. This is not known at this point. In any case, the will be a remaining unknown component of the self-gravity acceleration even after the launch of LISA. The current section will analyse the impact of such an unknown component. For the current analysis it has been assumed that this unknown component has a Gaussian distribution with a standard deviation of $1$ nm/s$^2$ per spacecraft axis. It has been assumed constant in the spacecraft-fixed frame during the mission time. This is an oversimplification and represents a conservative case until it can be better quantified how this acceleration changes over the mission time.
The results presented in Figure~\ref{fig:cartwheel-dispersion-withSG} and Table~\ref{tab:stabilityStatistics-withSG}  and show a substantial increase both in the corner angle deviation and in the arm length rate compared to those in section~\ref{subsec:stability}.

\begin{table}
\caption{Formation stability statistics for MIDA$=-20^\circ$ Range, Doppler case with added random self-gravity accelerations.}
\label{tab:stabilityStatistics-withSG}     
\begin{tabular}{lll}
\hline\noalign{\smallskip}
          &   Minimum    & Maximum \\
\noalign{\smallskip}\hline\noalign{\smallskip}
Overall Min./Max. Armlength [km]               &     2427721.9   &             2550716.5\\
$99 \%$ C.I. of Min./Max. Armlength [km]       &    2436408.7     &           2542065.1\\
Overall Min./Max. Armlength Rate [m/s]         &        -12.95    &                11.35\\
$99 \%$ C.I. of Min./Max. Armlength Rate [m/s] &        -11.04    &                10.16\\
Overall Min./Max. Corner Angle [deg]           &        58.207    &               61.817\\
$99 \%$ C.I. of Min./Max. Corner Angle [deg]   &        58.507    &               61.527\\
$99 \%$ C.I. Time spent beyond $60^\circ\pm 1.0^\circ$ [days] &   512    &\\
\noalign{\smallskip}\hline
\end{tabular}
\end{table}

\subsection{Cartwheel stability per component}
\label{subsec:stabilityPerComponent}
Finally, in this section the importance of the initial spacecraft dispersion along the individual axes is analysed. To this end, instead of using the dispersion matrices coming from the navigation analysis, as was described in section~\ref{subsec:navResults}, a fixed value along the different axes in the local orbital frame is assumed while the other axes covariances are set to zero. Additionally, a case is with a spherical position or velocity dispersion is run to compare the other cases to. 
The resulting statistics are summarized in Table~\ref{tab:stabilityStatistics-perAxis}.

\begin{table}
\caption{Statistics of different cases with pre-defined initial dispersion along individual axes.}
\label{tab:stabilityStatistics-perAxis}     
\begin{tabular}{lllll}
\hline\noalign{\smallskip}
Case    &   \multicolumn{2}{l}{$99 \%$ C.I. of}    & \multicolumn{2}{l}{$99 \%$ C.I. of} \\
        &   \multicolumn{2}{l}{Min./Max. Armlength Rate [m/s]}    & \multicolumn{2}{l}{Min./Max. Corner Angle [$^\circ$]} \\
\noalign{\smallskip}\hline\noalign{\smallskip}
200 km spherical ($1-\sigma$)	  &  -28.10 &                 	 21.75	&56.993&                   	63.035\\
200 km along-track ($1-\sigma$) &	-8.76 &                   	10.08	&58.965&                   	61.054\\
200 km cross-track ($1-\sigma$) &	-8.53 &                   	10.06	&58.984&                   	61.033\\
200 km ~radial ($1-\sigma$)	 &   -28.32 &                   	22.09	&57.007&                   	63.111\\
1 cm/s spherical ($1-\sigma$)  &	-12.83 &                   	10.96	&58.536&                   	61.484\\
1 cm/s along-track ($1-\sigma$)&	-12.81 &                   	10.99	&58.533&                   	61.479\\
1 cm/s cross-track ($1-\sigma$)&	-8.51 &                    	9.95	&58.988&                   	61.031\\
1 cm/s ~radial ($1-\sigma$)    &	-8.58 &                   	10.01	&58.981&                   	61.038\\
\noalign{\smallskip}\hline
\end{tabular}
\end{table}

From these results it is clear that the dominant contributions to the de-stabilization of the cartwheel are the ~radial position dispersion and the along-track velocity dispersion. These are the components which are associated with the semi-major axis. The cross-track component has the smallest contribution. Considering that ground-based Doppler observations mainly measure the along-track velocity component for LISA, this is a very favourable result. Added $\Delta$DOR measurements will therefore offer only a limited benefit.

To understand the dynamics a little better, also cases have been analysed where the initial dispersion was set to zero in all six state components and only self-gravity acceleration along individual axes has been added. The results for the self-gravity accelerations defined in the local orbital frame are shown in Table~\ref{tab:stabilityStatistics-SGperAxis}.

\begin{table}
\caption{Statistics of different cases with zero initial dispersion and constant self-gravity accelerations along individual axes in the local orbital frame.}
\label{tab:stabilityStatistics-SGperAxis}     
\begin{tabular}{lllll}
\hline\noalign{\smallskip}
Case    &   \multicolumn{2}{l}{$99 \%$ C.I. of}    & \multicolumn{2}{l}{$99 \%$ C.I. of} \\
        &   \multicolumn{2}{l}{Min./Max. Armlength Rate [m/s]}    & \multicolumn{2}{l}{Min./Max. Corner Angle [$^\circ$]} \\
\noalign{\smallskip}\hline\noalign{\smallskip}
1 nm/s$^2$ spherical ($1-\sigma$)  &	-103.49  &                    	93.20	 & 49.193 &                    	72.208 \\
1 nm/s$^2$ along-track ($1-\sigma$)  &	-102.70  &                    	93.57	 & 49.118 &                    	72.127 \\
1 nm/s$^2$ cross-track ($1-\sigma$)  &	-8.49  &                     	9.96	 & 58.988 &                    	61.031 \\
1 nm/s$^2$ ~radial ($1-\sigma$)  &  	-9.80  &                     	9.97	 & 58.847 &                    	61.153 \\
\noalign{\smallskip}\hline
\end{tabular}
\end{table}

It is clear from these results that again the main contribution to the cartwheel de-stabilization comes from the along-track acceleration component. In contrast to an initial along-track velocity dispersion, a constant along track acceleration has a much stronger impact on the formation stability. The radial and in particular the cross-track acceleration have a minor impact.
Since, however, the self-gravity accelerations are tied to the spacecraft-fixed frame, the analysis was repeated defining constant accelerations in the frame where:
\begin{itemize}
\item X – along the formation centre
\item Z – perpendicular to cartwheel plane
\item Y – completing the right-handed frame
\end{itemize}
Results are shown in Table~\ref{tab:stabilityStatistics-SGperSCAxis}. Because the spacecraft frame is rotating with a period of a year, there is no constant along-track acceleration. Therefore, the overall impact is milder than observed in the local orbital frame. The X and Y components show the strongest contribution to the cartwheel de-stabilization.

\begin{table}
	\caption{Statistics of different cases with zero initial dispersion and constant self-gravity accelerations along individual axes in the spacecraft-fixed frame.}
	\label{tab:stabilityStatistics-SGperSCAxis}     
	\begin{tabular}{lllll}
	\hline\noalign{\smallskip}
	Case    &   \multicolumn{2}{l}{$99 \%$ C.I. of}    & \multicolumn{2}{l}{$99 \%$ C.I. of} \\
			&   \multicolumn{2}{l}{Min./Max. Armlength Rate [m/s]}    & \multicolumn{2}{l}{Min./Max. Corner Angle [$^\circ$]} \\
	\noalign{\smallskip}\hline\noalign{\smallskip}
	1 nm/s$^2$ spherical ($1-\sigma$)  &	-10.49  &                    	10.10	 & 58.523 &                    	61.513 \\
	1 nm/s$^2$ x-s/c ($1-\sigma$)  &    	-9.26  &                     	9.97	 & 58.713 &                    	61.351 \\
	1 nm/s$^2$ y-s/c ($1-\sigma$)  &	    -9.69  &                     	9.96	 & 58.651 &                    	61.358 \\
	1 nm/s$^2$ z-s/c ($1-\sigma$)  &	    -9.61  &                     	9.96	 & 58.867 &                    	61.131 \\	
	\noalign{\smallskip}\hline
	\end{tabular}
	\end{table}

%%%%%%%%%%%%%%%%%%%%%%%%%%%%%%%%%%%%%%%%%%%%%%%%%%%%%%%%%%%%%%%%%%%
\section{Conclusions}
\label{sec:conclusions}
%%%%%%%%%%%%%%%%%%%%%%%%%%%%%%%%%%%%%%%%%%%%%%%%%%%%%%%%%%%%%%%%%%%

After a review of analytic models for the LISA cartwheel formation, a fully numerical optimisation analysis of both the transfer and science phase have been presented. The interdependency of both mission phases via the cartwheel clocking angle has been taken into account. The SEP transfer \dv~varies depending on the launch month in 2034. But for the sizing month an allocation of $1092$ m/s per spacecraft is sufficient with the assumptions taken. For the science orbit, with the assumed arm length of $2.5\times 10^6$ km and MIDA=$-20^\circ$, a corner angle variation of close to $60^\circ \pm 1.0^\circ$ during 10 years is feasible.

The expected cartwheel insertion accuracy has been estimated in a full navigation analysis where the most important sources of error have been taken into account. An accuracy of the order of $\mathcal{O}(10 \text{ km}, 5 \text{ mm/s})$ (along-track and radial) and $\mathcal{O}(100 \text{ km}, 50 \text{ mm/s})$ (cross-track) are achievable with Range/Doppler and several weeks of insertion sequence using 18 m/s per spacecraft. The radial position error and along-track velocity error are the driving ones for the stability of the subsequent science cartwheel orbit.

With the obtained insertion dispersion, a Monte-Carlo analysis has been conducted to analyse the impact on the corner angle variations during 10 years of science phase. If the self-gravity accelerations are perfectly known at the time of cartwheel insertion, the expected deterioration of corner angle variations is about $60^\circ \pm 1.1^\circ$ at $99\%$ C.L.. The arm length rate is hardly impacted. However, if there is a remaining unknown constant component of the self-gravity accelerations of the order of 1 nm/s$^2$, the impact on the corner angle variations and arm length rate is significant. This underlines the importance of a good characterisation of the self-gravity accelerations prior to launch.

In the frame of the currently ongoing Phase A, the following future Mission Analysis work is envisioned:
\begin{itemize}
	\item A better objective function for the cartwheel orbit optimisation: not applying strict constraints on the corner angles and arm length rates, but rather maximising the mission time spent within the allowed bands.
	\item Cartwheel orbit optimisation including spacecraft self-gravity.
	\item Relaxation of the maximum Earth distance constraint taking into account the actual achievable data rate as a function of the Earth distance.
	\item Implementation of a variable thrust model for the transfer optimisation. Instead of a constant thrust value, variations of the thrust as a function of Sun distance and SAA shall be taken into account.
\end{itemize}

\begin{acknowledgements}
The authors would like to thank Michael Khan for his LISA Mission Analysis work during the ESA CDF study which created the foundation of the work presented in this paper. Moreover, the authors acknowledge the fruitful and stimulating collaboration between ESA and the two prime contractors, Airbus and Thales.
\end{acknowledgements}

\section*{Conflict of interest}
The authors declare that they have no conflict of interest.



\end{document}